\documentclass[twocolumn]{aastex631}

\shorttitle{SN 2011fh}
\shortauthors{Pessi et al.}
\graphicspath{{./}{}}

\begin{document}

\title{Unveiling the Nature of SN 2011fh: a Young and Massive Star Gives Rise to a Luminous SN 2009ip-like Event}

\author[0000-0001-6540-0767]{Thallis Pessi}
\altaffiliation{E-mail: thallis.pessi@mail.udp.cl}
\affiliation{Núcleo de Astronomía, Universidad Diego Portales, Av. Ejército 441, Santiago, Chile}

\author[0000-0003-1072-2712]{Jose L. Prieto}
\affiliation{Núcleo de Astronomía, Universidad Diego Portales, Av. Ejército 441, Santiago, Chile}
\affiliation{Millennium Institute of Astrophysics, Santiago, Chile}

\author{Berto Monard}
\affiliation{Klein Karoo Observatory, Calitzdorp, 
South Africa}

\author[0000-0001-6017-2961]{Christopher S. Kochanek}
\affiliation{Department of Astronomy, The Ohio State University, 140 West 18th Avenue, Columbus, OH 43210, USA}
\affiliation{Center for Cosmology and Astroparticle Physics, The Ohio State University, 191 W. Woodruff Avenue, Columbus, OH 43210, USA}

\author{Greg Bock}
\affiliation{Runaway Bay, Gold Coast, Queensland, Australia}

\author{Andrew J. Drake}
\affiliation{Division of Physics, Mathematics, and Astronomy,
California Institute of Technology, Pasadena,
CA 91125, USA}

\author[0000-0003-2238-1572]{Ori D. Fox}
\affiliation{Space Telescope Science Institute, 3700 San Martin Drive, Baltimore, MD 21218, USA}

\author{Stuart Parker}
\affiliation{Parkdale Observatory, 225 Warren Road, RDl Oxford, Canterbury 7495, New Zealand}

\author[0000-0002-0504-4323]{Heloise F. Stevance}
\affiliation{Department of Physics, The University of Auckland, Private Bag 92019, Auckland, New Zealand}

\begin{abstract}

In recent years, many Type IIn supernovae have been found to share striking similarities with the peculiar SN 2009ip, whose true nature is still under debate. Here, we present 10 years of observations of SN 2011fh, an interacting transient with spectroscopic and photometric similarities to SN 2009ip. SN 2011fh had a M$_r \sim -16$ mag brightening event, followed by a brighter M$_r \sim -18$ mag luminous outburst in August 2011. The spectra of SN 2011fh are dominated by narrow to intermediate Balmer emission lines throughout its evolution, with P Cygni profiles indicating fast-moving material at  $\sim 6400 \ \textrm{km s}^{-1}$. HST/WFC3 observations from October of 2016 revealed a bright source with M$_{F814W} \approx -13.3$ mag, indicating that we are seeing the ongoing interaction of the ejecta with the circumstellar material or that the star might be going through an eruptive phase five years after the luminous outburst of 2011. Using HST photometry of the stellar cluster around SN 2011fh, we estimated an age of $\sim 4.5$ Myr for the progenitor, which implies a stellar mass of $\sim 60$ M$_\odot$, using single-star evolution models, or a mass range of $35 - 80 \ \textrm{M}_\odot$, considering a binary system. 
We also show that the progenitor of SN 2011fh exceeded the classical Eddington limit by a large factor in the months preceding the luminous outburst of 2011, suggesting strong super-Eddington winds as a possible mechanism for the observed mass-loss. These findings favor an energetic outburst in a young and massive star, possibly a luminous blue variable.

\end{abstract}


\section{Introduction} \label{sec:intro}

Most stars with a zero age main sequence (ZAMS) mass higher than 8 M$_\odot$ are thought to end their lives as a core-collapse supernova (CCSN). An important fraction of these explosions happen inside a very dense environment, formed by the material ejected by the star in its previous mass-loss history. If the circumstellar material (CSM) is sufficiently dense and dominated by hydrogen, 
hydrogen emission lines with multiple velocity components appear, leading to a Type IIn supernova \cite[SN,][]{1990MNRAS.244..269S, 1997ARA&A..35..309F,2021MNRAS.506.4715R}. If, on the other hand, the CSM is helium-rich, a Type Ibn SN is observed \citep{2007Natur.447..829P, 2008MNRAS.389..113P}. Recently, Type Icn SNe have been discovered, with narrow C and O lines, and no H and He lines \citep{2021TNSAN..76....1G, 2021arXiv210807278F}. 

These events are sometimes referred to as interacting SNe, since the characteristic emission lines in their spectra come from the ionization of the CSM, by emission from the SN shock \citep{2017hsn..book..403S, Fraser2020}. Other observational properties of these transients include a relatively blue continuum, very diverse and prolonged light curves \citep{2015A&A...580A.131T, 2020A&A...637A..73N} and, for a fraction of interacting SNe, 
enhanced X-ray \citep{2000MNRAS.319.1154F, 2014ApJ...780..184K, 2012ApJ...755..110C, 2012ApJ...750L...2C} or radio emission \citep{1993ApJ...419L..69V, 2002ApJ...572..932P, 2002ApJ...581..396W, 2009ApJ...690.1839C}. 

The fact that we observe interacting SNe indicates that intense mass loss activity can occur in the years preceding the collapse of the progenitor star, and we still do not have a good physical understanding of the mechanisms involved \citep[see the review by][]{2014ARA&A..52..487S}.
Standard stellar evolution models do not account for the amount of mass-loss needed to explain these transients, limiting our understanding of their progenitors \citep[][]{2012ARA&A..50..107L}.
In fact, direct observations of the progenitor stars of Type IIn/Ibn SNe have been elusive, with only a few being confirmed \citep{2007ApJ...656..372G, 2009Natur.458..865G, 2010AJ....139.1451S, 2011MNRAS.415..773S, 2011ApJ...732...32F}. 

The progenitors of interacting SNe can be diverse, which is highlighted by the extreme diversity in their observational properties \citep{2015A&A...580A.131T}. The progenitor of the Type IIn SN 2005gl was a  massive luminous blue variable (LBV)-like star \citep{2009Natur.458..865G}, so at least some of these events are generated by LBV-like massive stars. Analysis of the environments of these explosions, however, show that interacting SNe happen in very similar locations to Type II SNe \citep[][]{2012MNRAS.424.1372A, 2017A&A...597A..92K}, indicating that they might have overlapping mass ranges, and are not confined to very massive stars.  

Some massive stars can go through very unstable mass-loss phases, ejecting large amounts of matter in energetic eruptions and explosions \citep{2014ARA&A..52..487S}.
One well-known example is the massive Galactic LBV Eta Carinae, which went through a major eruption event (the Great Eruption) in the 19th century reaching an absolute magnitude of M$_V \sim -14$ and with properties similar to Type IIn SNe \citep{2012Natur.482..375R, 2014ApJ...787L...8P, 2018MNRAS.480.1466S, 2018MNRAS.480.1457S}. Estimates show that the star produced $\sim 10^{50}$ ergs of kinetic energy and ejected $10-20$ M$_\odot$ of material \citep{2003AJ....125.1458S, 2006ApJ...644.1151S}. The main mechanism that caused the Great Eruption of Eta Carinae is still under debate, although recent observations of its light echoes point to a possible merger in a triple system \citep{2018MNRAS.480.1466S, 2021MNRAS.503.4276H}.  

Some of these giant stellar eruptions happen inside a very dense medium, and produce spectroscopically and photometrically similar transients to interacting SNe. Although being a non-terminal event, in the sense that the progenitor star is still alive afterwards, these outbursts also interact with the CSM. Because of these similarities, some of these outbursts are mistakenly classified as SNe, originating the term `supernova impostors'. A famous impostor associated with a massive star progenitor is SN 1997bs, which had a peak at M$_V \simeq$ M$_R \simeq -14$, a very slow decline in its light curve, and a spectrum dominated by strong narrow H$\alpha$ in emission \citep{2000PASP..112.1532V}, although there is presently no evidence that the star survived \citep{2015MNRAS.452.2195A}. Recently, \citet[][]{2019A&A...630A..75P} alternatively proposed SN 1997bs to be a stellar merger. Another impostor, SN 2000ch, is thought to have a progenitor similar to Eta Carinae and at its peak of M$_V \simeq -12.7$ it showed a full width at half maximum (FWHM) of the Balmer lines up to $\sim 1550$ km s$^{-1}$ \citep{2000IAUC.7421....3F, 2004PASP..116..326W}. 
It is important to note that SN impostors can also arise from lower mass stars, such as SN 2008S and similar events \citep{2012ApJ...758..142K}, whose nature and that of other transients like it is still under debate \citep[they might have been real CCSNe, see e.g.,][]{2011ApJ...741...37K,2016MNRAS.460.1645A, 2018MNRAS.480.3424C, 2021arXiv210805087C}. These transients have dust-enshrouded progenitors and ZAMS masses of $\sim 6-10$ M$_\odot$ \citep{2008ApJ...681L...9P,2009ApJ...705.1364T}.

A remarkable example of the confusion between real CCSNe and impostors is SN 2009ip. When discovered in 2009 \citep{2009CBET.1928....1M} it was classified as a Type IIn SN, reaching an absolute magnitude of M$_R \sim -14.5$.  One year later, another bright eruption with similar luminosity was detected \citep{2013ApJ...767....1P}, indicating that the progenitor was still alive and did not experience a core-collapse event in 2009. These eruptions had, in fact, many similarities with LBVs, showing velocities of a few hundred km s$^{-1}$, although faster material with thousands of km s$^{-1}$ was also observed. Analysis of pre-discovery images suggested that the progenitor star of SN 2009ip was a very massive star with 50–80 M$_\odot$, showing LBV-like variability in the years before the 2009 eruption \citep{2010AJ....139.1451S, 2011ApJ...732...32F}. 

In 2012, SN 2009ip had two bright explosive events, reaching M$_V \sim -14.5$ in August (usually labeled Event A), and M$_R \sim -18$ in October (called Event B). Several authors have discussed the nature of the luminous event \citep{2013MNRAS.430.1801M, 2013ApJ...767....1P, 10.1093/mnras/stt813}. \cite{2013MNRAS.430.1801M} claimed that the high velocities detected in the outburst is evidence for a true CCSN during an LBV-like activity. \cite{2013ApJ...767....1P}, however, argued that such high velocities were also observed during the 2011 event, which was clearly not a true SN. This scenario was also supported by \cite{10.1093/mnras/stt813}, who showed that there was no signal of nebular emission from nucleosynthesis products after the 2012B eruption, although nebular emission could be hidden by the continuing interaction with CSM. 

In the subsequent years, new analyses and a number of new observations have contributed to the discussion of whether the 2012 eruptions of SN 2009ip were a terminal event or not \citep{2013ApJ...763L..27P,2014AJ....147...23L, 2014ApJ...780...21M, 2014MNRAS.438.1191S, 2014ApJ...787..163G, 2014MNRAS.442.1166M, 2015ApJ...803L..26M, 2013MNRAS.433.1312F, 2015MNRAS.453.3886F, 2015ATel.8417....1T, 2016MNRAS.463.2904S, 2017MNRAS.469.1559G}. A number of newly detected transients also presented very similar behavior to SN 2009ip. SN 2010mc,  classified  as  a  Type  IIn  SN \citep{2013Natur.494...65O},  shows a nearly identical light curve to 2009ip, with precursor variability, a decline after peak consistent with $^{56}$Co decay, and late-time flattening in the light curve due to CSM interaction. \cite{2014MNRAS.438.1191S} claimed that SN 2010mc was, in fact, a core-collapse event, arguing that the missing nebular phase is explained by CSM obscuration, and that a non-terminal event with energies of $\sim 10^{50}$ erg is implausible.
SN 2016bdu \citep{10.1093/mnras/stx2668} also had spectroscopic features of Type IIn  SNe, with several similarities to SN 2009ip, including the remarkably similar light curve for its two explosive events. Pre-discovery images also showed progenitor variability similar to that of SN 2009ip.
SN 2018cnf \citep{2019A&A...628A..93P} also had a sequence of two bright outbursts, after years of explosive variability with eruptions reaching M$_r \sim -14.7$. \cite{2019A&A...628A..93P} proposed a massive hypergiant or an LBV star as a plausible progenitor for SN 2018cnf. 
More recently, the event AT2016jbu/Gaia16cfr showed a variable behavior followed by two bright outbursts \citep{2018MNRAS.473.4805K,2021arXiv210209572B, 2021arXiv210209576B}. The availability of pre-discovery images from the Hubble Space Telescope (HST) allowed \cite{2021arXiv210209572B, 2021arXiv210209576B} to estimate that the progenitor star had a relatively low mass of $\sim 20$ M$_\odot$.
Other notable SN 2009ip-like transients might include SN 2015bh \citep{2016MNRAS.463.3894E, 2017A&A...599A.129T, 2018A&A...617A.115B}, SNhunt151 \citep{2018MNRAS.475.2614E}, LSQ13zm \citep[][]{2016MNRAS.459.1039T}, and SN 2013gc \citep[][]{2019MNRAS.482.2750R}.

The growing class of SN 2009ip-like transients show several observational similarities: i) erratic pre-discovery variability, sometimes including LBV-like outburts; ii) two luminous eruptive events, with the second peak reaching $\sim -18$ mag; and iii) the spectroscopic properties of Type IIn SNe, including multiple  velocity components observed in the Balmer lines. 

Here, we present observations of SN 2011fh, which share many similarities to SN 2009ip. Discovered by amateur astronomer Berto Monard on 2011 August 24 from South Africa, SN 2011fh was classified as a Type IIn SN at peak by \cite{2011CBET.2799....2P}. The source is located in the outskirts of the nearby galaxy NGC 4806. We describe our 10 years of observations of SN 2011fh in Section \ref{sec:obs}. Section \ref{sec:resul} describes the host galaxy (Section \ref{sec:host_gal}), the photometric (Section \ref{sec:phot_evol}) and spectroscopic (Section \ref{sec:spec_evol}) evolution, and the environment where SN 2011fh is located (Section \ref{sec:host}). We discuss our analysis in Section \ref{sec:dis}, and present a final summary of our conclusions in Section \ref{sec:concl}. 

\section{Observations} \label{sec:obs}

SN 2011fh was discovered by amateur astronomer Berto Monard on 2011 August 24 UT\footnote{UT dates are used throughout this paper.} using unfiltered images from the Klein Karoo Observatory in South Africa \citep{2011CBET.2799....1M}, at $\sim 14.5$ mag. The source was classified as a Type IIn SN at peak by \cite{2011CBET.2799....2P}, using observations obtained with the du Pont 2.5-m telescope and the Wide Field reimaging CCD camera (WFCCD) at the Las Campanas Observatory (LCO) on 2011 August 28. A series of spectroscopic observations were obtained after peak brightness using the Apache Point Observatory (APO) 3.5 m telescope, the du Pont 2.5 m telescope, and the Magellan Baade and Magellan Clay 6.5 m telescopes. Photometric measurements were obtained by the Carnegie  Supernova  Project-II \citep[CSP2,][]{ 2006PASP..118....2H, 2019PASP..131a4001P, 2020A&A...639A.103S} using the Swope 1-m and du Pont 2.5-m telescopes, and archival data were retrieved from the Catalina Real-Time Transient Survey \citep[CRTS,][]{2009ApJ...696..870D} and amateur astronomers. We also analyze archival images obtained by the Spitzer Space Telescope (SST) and the Hubble Space Telescope (HST), as well as spectroscopic observations made with the Multi-Unit Spectroscopic Explorer (MUSE) instrument on the Very Large Telescope (VLT).

\begin{figure*}[t]
\plotone{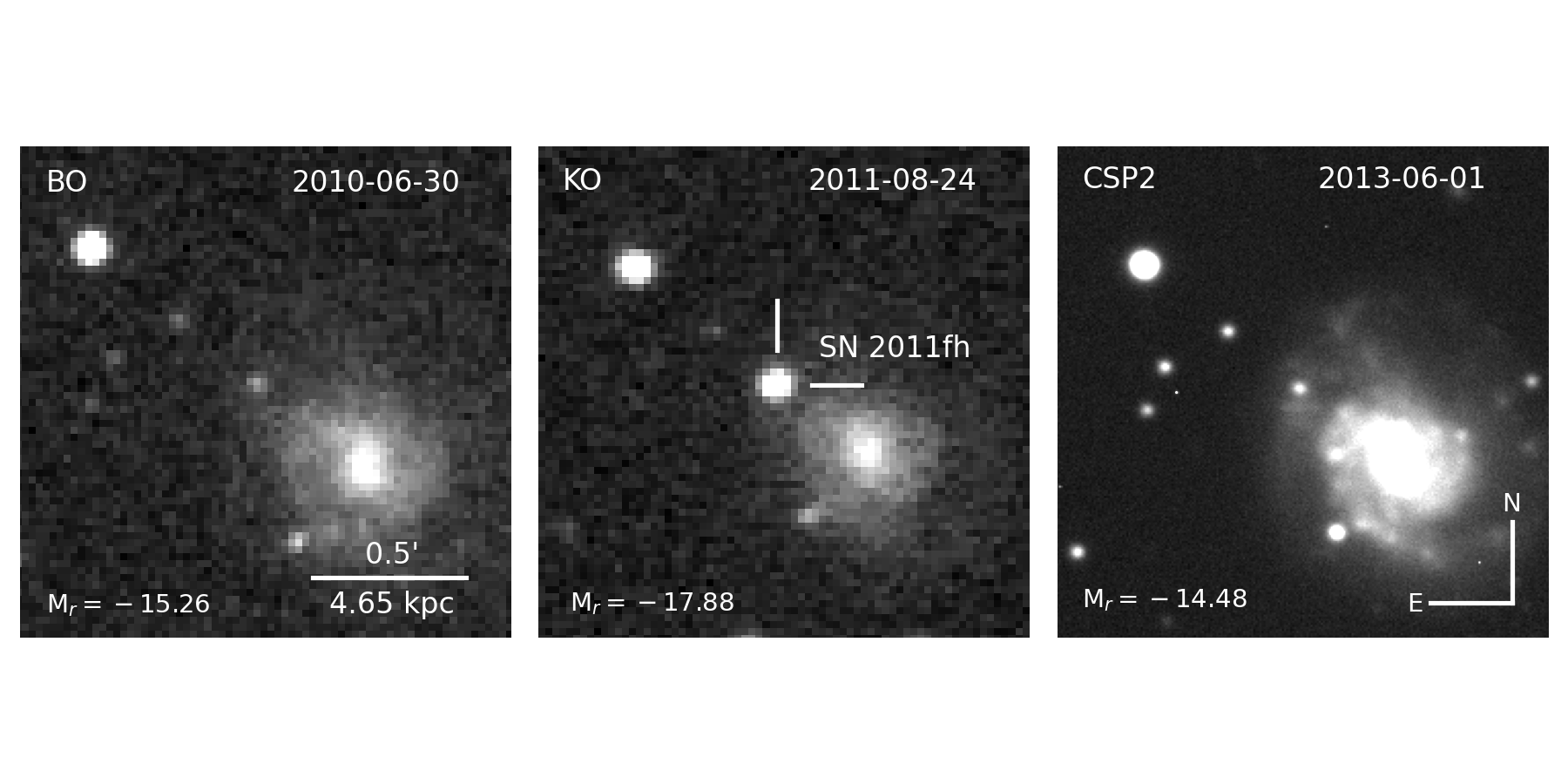}
\caption{Unfiltered images of the galaxy NGC~4806 and SN~2011fh obtained by Berto Monard (left and middle panel) and a r-band CSP2 image at a later phase (right panel). SN 2011fh reached a peak absolute magnitude of M$_r = -17.88$ on 2011 August 24, as it is shown in the middle panel. The $0.5\arcmin$ horizontal bar corresponds to 4.65 kpc at the distance of NGC 4806. The panels also show the UT date when the images were taken, and the absolute $r$-band magnitude of SN 2011fh at that time. 
\label{fig:fig1}}
\end{figure*}

\subsection{Optical and NIR Photometry}

Archival unfiltered images obtained by amateur astronomers and CRTS span from 2007 to 2013 (Figure \ref{fig:fig1}). 
Berto Monard obtained 27 unfiltered images between 2009 January 06 and 2013 January 22, using Meade 12- and 14-inch f/8 telescopes, and an SBIG ST-8 CCD with a pixel scale of 1.6 arcsec/pixel.
Five images were obtained by Stu Parker using a Celestron 14-inch f6.3 telescope with a pixel scale of 1.1 arcsec/pixel, and a Celestron 11-inch f6.3 telescope with a pixel scale of 1.6 arcsec/pixel, both using an SBIG ST-10 3 CCD camera, between 2011 June 16 and August 09.
Three images were also obtained by Greg Bock using a Meade 14-inch telescope, and an SBIG ST-10 Dual CCD Camera, with a pixel scale of 1.1 arcsec/pixel, between 2011 June 21 and 2012 February 17. 
CRTS unfiltered images were all obtained by the Siding Springs Survey 0.5m telescope with a pixel scale of 1.8 arcsec/pixel. 

\begin{figure*}[t]
\epsscale{1.2}
\centerline{\includegraphics[scale=.5]{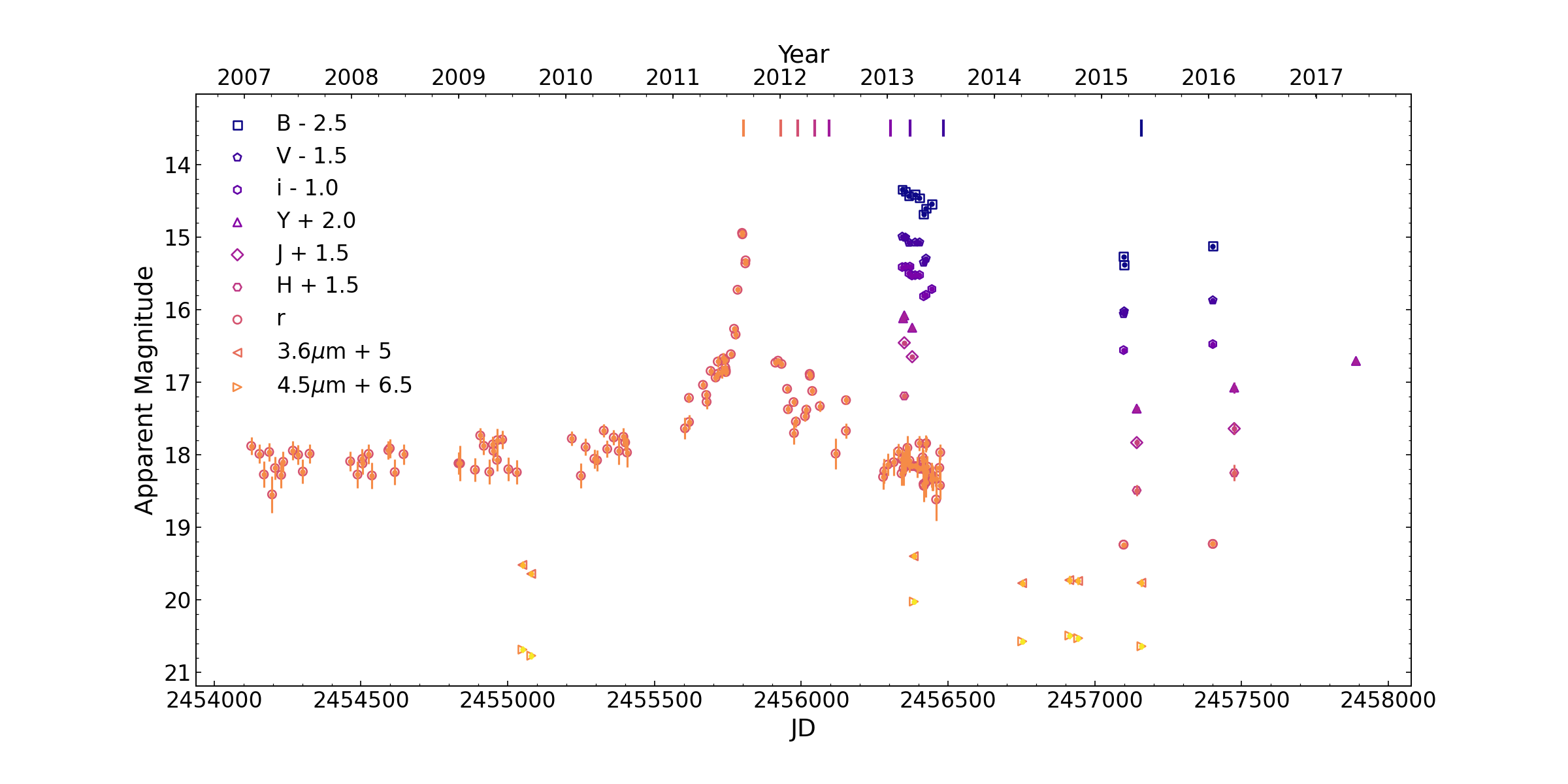}}
\caption{The complete photometry of SN 2011fh, including data from amateur astronomers, CRTS, CSP2, and the Spitzer Space Telescope. Unfiltered images from amateur astronomers and the CRTS were calibrated to the SDSS $r$-band magnitudes and span from 2007 January 25 to 2013 June 29. SN 2011fh was observed by the CSP2 from 2013 February 20 to 2017 May 16 using the $BVriYJH$ filters. The Spitzer observations include early detections on 2009 August 6 and September 5, and five additional observations from 2013 March 31 to 2015 May 16. Spectra were obtained at the epochs marked as vertical lines at the top of the figure.  \label{fig:fig2}} 

\end{figure*}

The unfiltered images were reduced using standard procedures, including bias and dark subtraction, and flat-fielding. The photometric measurements were made using point spread function fitting (PSF) through {\tt\string DOPHOT} \citep{1993PASP..105.1342S, 2012AJ....143...70A}.
The measured fluxes were calibrated to CSP2 $r$-band magnitudes using stars in the field. The SDSS $r$-band was chosen because the emission is strongly dominated by H$\alpha$, as it is usual in transients arising from the interaction with hydrogen rich CSM, and also because this band is roughly at the midpoint of the wavelengths spanned by the unfiltered response function. The photometric measurements are presented in Table \ref{tab:r-crts} and the $r$-band light curve of SN 2011fh is shown in Figure \ref{fig:fig2}. 

After peak, $BVri$ images of SN 2011fh were obtained by the Carnegie Supernova Project-II (CSP2), with observations spanning from 2013 February 20 to 2016 January 14 using the direct imaging cameras of the du Pont 2.5 m telescope and the Henrietta Swope 1.0 m telescope (Figure \ref{fig:fig1}) at LCO. Near-infrared (NIR) $YJH$ images \citep{2017AJ....154..211K}, were taken between 2013 February 23 and 2017 May 16 with RetroCam on the du Pont telescope.

The CSP2 images were reduced using standard techniques in IRAF \citep{2005ASPC..339...50F, 2006PASP..118....2H, 2020A&A...639A.103S}. PSF photometry of SN 2011fh was extracted using {\tt\string DOPHOT}, and the calibration used CSP2 photometry of standard stars in the field. The optical $BVri$ photometry is reported in Table \ref{tab:BVri-csp2} and the NIR $YJH$ photometry in Table \ref{tab:YJH-csp2}. The CSP2 light curves of SN 2011fh are shown in Figure \ref{fig:fig2}.

\subsection{Spectroscopy}

SN 2011fh was classified as a Type IIn SN $\sim 3$ days after peak brightness using a spectrum obtained by the du Pont 2.5 m Telescope with the WFCCD camera. The spectrum was obtained using a $1.7\arcsec$ slit with a FWHM resolution of $\sim 8$\AA. Three other spectra obtained with the same instrument and similar configurations were taken on 
2012 June 15, 2013 October 7, and on 2013 January 12. Four optical spectra were also obtained with the APO 3.5 m Telescope on 2012 January 01, 2012 February 29, 2012 April 29, and on 2013 March 18, using the Dual Imaging Spectrograph (DIS), with a $1.5\arcsec$ slit and a FWHM resolution of $\sim 7$\AA. A high-resolution spectrum ($\Delta \lambda = 0.2 $ \AA) was obtained with the Magellan Clay 6.5m Telescope using the Magellan Inamori Kyocera Echelle \citep[MIKE,][]{2003SPIE.4841.1694B} spectrograph with a $1\arcsec$ slit on 2011 August 31. An additional near-infrared (NIR) spectrum was obtained with the Folded port InfraRed Echellette \citep[FIRE, ][]{2013PASP..125..270S} spectrograph on the Magellan Baade 6.5m Telescope, using a $0.6\arcsec$ slit, and the high-throughput prism mode, which gives continuous coverage from 0.8-2.5 micron and resolutions of R $\sim 500$, 450, and 300 in the $JHK$ bands, respectively.

The low resolution optical spectra were reduced using standard techniques in IRAF {\tt twodspec} and {\tt onedspec} packages. We also applied a constant flux calibration correction to the spectra to match the synthetic $r$-band photometry to the light curve. The MIKE high-resolution spectrum was reduced using the CarPy reduction pipeline. The NIR FIRE observations were obtained using an ABBA nodding sequence and a bright A0V star close to the target in the sky was used as a flux and telluric standard. The spectrum was reduced using the IDL pipeline {\tt\string FIREHOSE} \citep[][]{2013PASP..125..270S}. 

\begin{figure*}[t!]
\epsscale{1.3}
\centerline{\includegraphics[scale=.5]{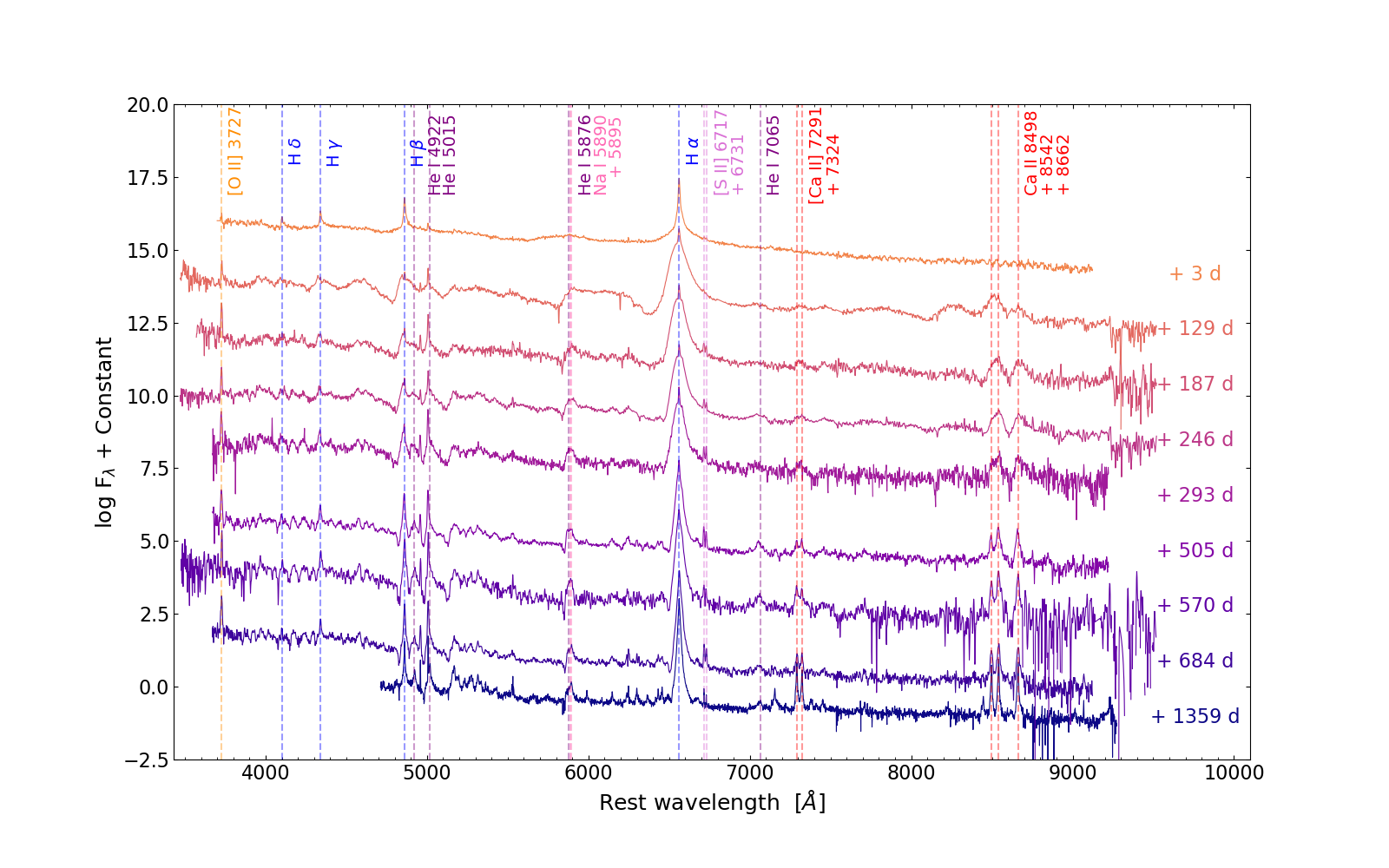}}
\caption{The spectral evolution of SN 2011fh from 3 (top) to 1359 (botton) days after peak brightness. Wavelength is in the rest frame, and we show log$_{10} \textrm{F}_\lambda$ plus an offset. The phase is relative to maximum brightness in $r$-band. Prominent lines of H, He I, Ca II, Na I, [S II], and [O II] are highlighted with dashed lines. The spectra at phases +3 d, +293 d, +505 d, and +684 d were obtained by the du Pont 2.5 m + WFCCD, and the spectra at phases +129 d, +187 d, +246 d, and +570 d by the APO 3.5 m + DIS instrument (Table \ref{tab:spec}). The last spectrum, at phase +1359 d, was extracted from VLT + MUSE observations. All the spectra presented in this figure are available in WISeREP \citep[\url{https://www.wiserep.org/},][]{2012PASP..124..668Y}. \label{fig:fig_spec}}
\end{figure*}

\begin{figure*}[t!]
\epsscale{1.3}
\centerline{\includegraphics[scale=.5]{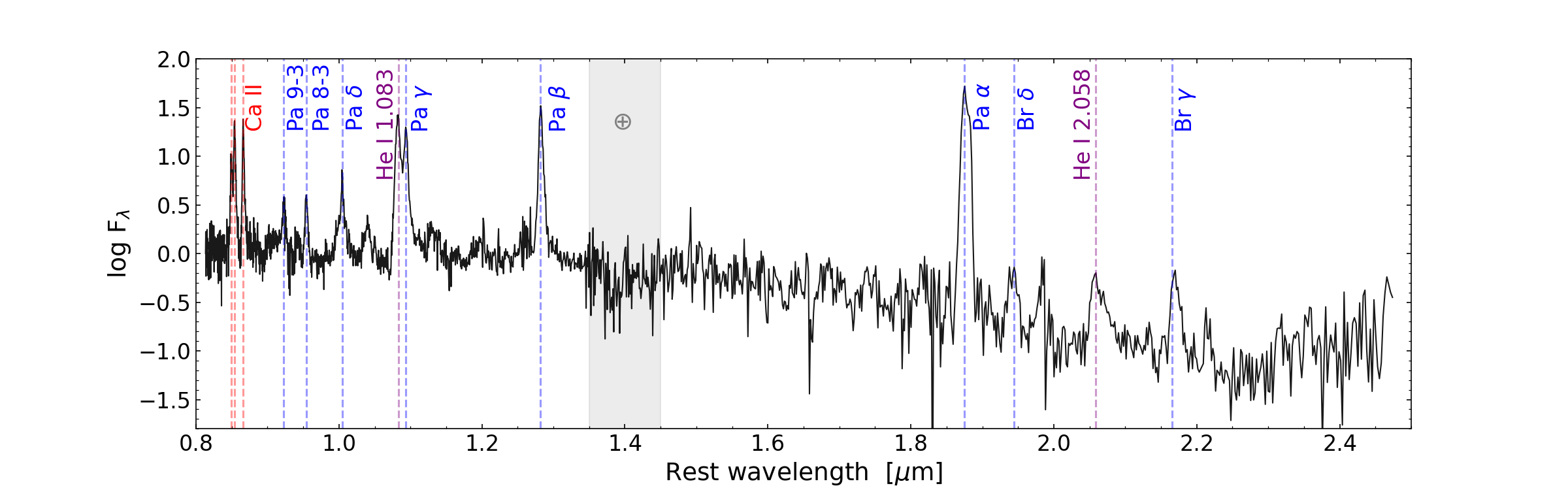}}
\caption{Rest frame FIRE NIR spectrum of SN 2011fh obtained on 2013 April 18, 600 days after peak brightness. The spectrum is dominated by emission lines of hydrogen, with prominent Paschen series (Pa$\alpha$, Pa $\beta$, Pa $\gamma$, Pa $\delta$, Pa $8-3$, and Pa $9-3$) and Brackett series (Br $\gamma$ and Br $\delta$) lines highlighted with by the blue dashed lines. Lines of He I at 1.083 and 2.058 $\mu$m are marked by purple lines. The Ca II $\lambda \lambda 8498, 8542, 8662$ triplet is marked by orange dashed lines. The gray band marks the region of the strong telluric absorption.  \label{fig:spec_fire}}
\end{figure*}

The spectra are summarized in Table \ref{tab:spec}, including the dates, instruments and configurations used for the observations. The spectral evolution of SN 2011fh is shown in Figure \ref{fig:fig_spec}, and the NIR FIRE spectrum is shown in Figure \ref{fig:spec_fire}.

\begin{deluxetable*}{ccclccc}
\tablecaption{Spectroscopic observations of SN 2011fh \label{tab:spec}}
\tablewidth{0pt}
\tablehead{
\colhead{Phase} &
\colhead{UT Date} & \colhead{HJD} & \colhead{Telecscope/Instrument} &
\colhead{Slit Width} & \colhead{Resolution} & \colhead{Wavelength Range}\\
\colhead{(days)} & \colhead{} & \colhead{} & \colhead{} &
\colhead{(arcsec)} & \colhead{(Angstrom) } & \colhead{(Angstrom)} 
}
\decimalcolnumbers
\startdata
+3 & 2011-08-30   & 2455803.479    &    du Pont 2.5m/WFCCD    &   1.7 & 8 & 3700 - 9120\\
+4 & 2011-08-31 & 2455805.479 & Magellan Clay/MIKE & 1.0 & 0.2 & 3500 - 9000 \\
+129 & 2012-01-02 & 2455928.983 & APO 3.5m/DIS & 1.5 & 7 & 3470 - 9520\\
+187 & 2012-02-29  &  2455986.895   &     APO 3.5m/DIS         &          1.5 & 7 & 3570 - 9520\\
+246 & 2012-04-29  &  2456046.740    &    APO 3.5m/DIS        &           1.5 & 7 & 3470 - 9520\\
+293 & 2012-06-15 &    2456093.612   &       du Pont 2.5m/WFCCD   &    1.7 & 8 & 3470 - 9520\\
+505 & 2013-01-12  &   2456304.846  &      du Pont 2.5m/WFCCD  &     1.7 & 8 & 3670 - 9225\\
+570 & 2013-03-18 &  2456369.840    &    APO 3.5m/DIS         &          1.5 & 7 & 3480 - 9520\\
+600 & 2013-04-18 &  2456400.507    &    Magellan Baade/FIRE         &         0.6 & 24/36/73 & 8130 - 24740 \\
+684 & 2013-07-10   &  2456484.465  &      du Pont 2.5m/WFCCD   &    1.7 & 8 & 3670 - 9120\\
+1359 & 2015-05-17   &  2457159.037  &     VLT/MUSE   &  0.8   & 2.3 & 4710 - 9270 \\
\enddata

\tablecomments{Since the VLT/MUSE spectrum comes from integral field data, $0.8\arcsec$ is the radius of a circular aperture around the source in the datacube, not the slit width.}

\end{deluxetable*}

\vspace{1cm}

\subsection{Spitzer Imaging}

After SN 2011fh was discovered, Spitzer/IRAC \citep{2004ApJS..154...10F} 3.6- and 4.5-micron archival images from 2009 August and September (Program 6007; PI: K. Sheth) revealed a mid-infrared  source at the position of the SN \citep{2011CBET.2799....1M}. Five more images using the 3.6 and 4.5-micron channels were taken from 2013 to 2015 (Programs 90137, 10139, 11053, PI: O. Fox). We downloaded the calibrated PBCD images from the Spitzer Heritage Archive\footnote{\url{https://sha.ipac.caltech.edu/applications/Spitzer/SHA/}} and performed PSF photometry using DOPHOT. The result of our photometry is reported in Table \ref{tab:spit}. The 3.6- and 4.6-micron evolution of SN 2011fh is also shown in Figure \ref{fig:fig2}. {The Spitzer/IRAC photometry of SN 2011fh 586 days after peak was also presented by \citet{2019ApJS..241...38S}, and is fairly consistent with our results}. \\

\vspace{1cm}

\subsection{HST Imaging} \label{sec:hst_imaging}

\begin{figure}[h]
\centerline{\includegraphics[scale=.35]{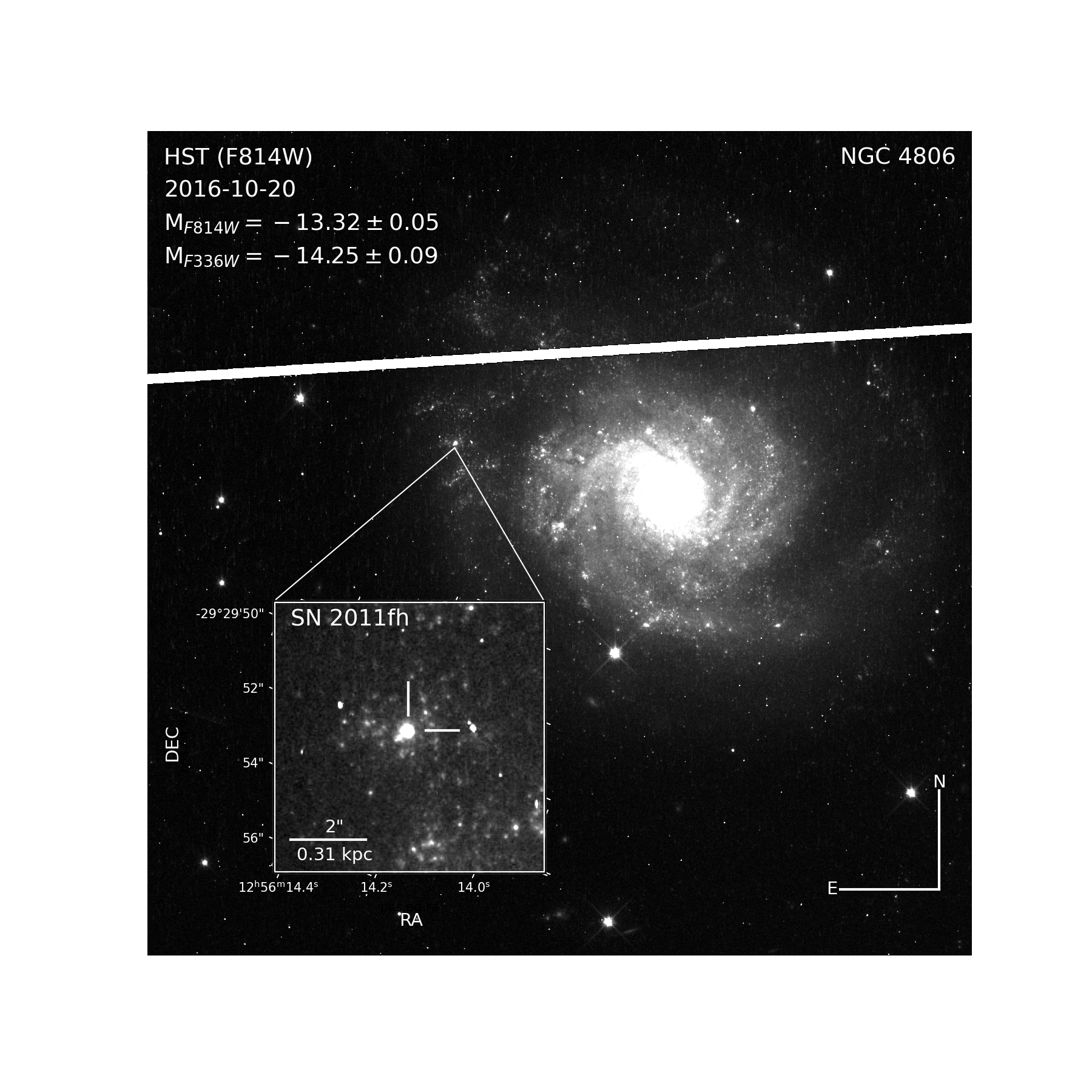}}
\caption{HST F814W image of NGC 4806 from 2016 October 20. The inset shows the region around SN 2011fh. The $2\arcsec$ horizontal bar at the inset corresponds to 0.31 kpc at the position of SN 2011fh. SN 2011fh has absolute magnitudes of M$_{F814W} = -13.32 \pm 0.05$ mag and M$_{F336W} = -14.25 \pm 0.09$ mag.  \label{fig:fig_hst}}
\end{figure}

SN 2011fh was observed by HST on 2016 October 20 (proposal ID 14149; PI: A.V. Filippenko) using the Wide Field Camera 3 (WFC3) Ultraviolet-Visible (UVIS) channel, and the F336W and F814W filters with exposure times of 710~s and 780~s, respectively. The data were retrieved from the Mikulski Archive for Space Telescopes (MAST). PSF photometry on the calibrated {\tt flt} images was performed with {\tt\string DOLPHOT} \citep{2016ascl.soft08013D}, using the drizzled F814W image as a reference for point source detection and astrometry. We selected point sources using the {\tt\string DOLPHOT} parameter criteria of \cite{2014ApJS..215....9W}: $sharpness^2 \leq 0.15$, $crowding \leq 1.0$, and $signal-to-noise \geq 5$. The F814W image is shown in Figure \ref{fig:fig_hst} with an inset zooming on the region around SN 2011fh. We measured the WFC3 Vega magnitudes of m(F336W) $ = 18.956 \pm 0.006$ mag and m(F814W) $ = 19.374 \pm 0.004$ mag for the SN.

\subsection{MUSE Spectroscopy}

\begin{figure*}[t]
\epsscale{1.3}
\centerline{\includegraphics[scale=.5]{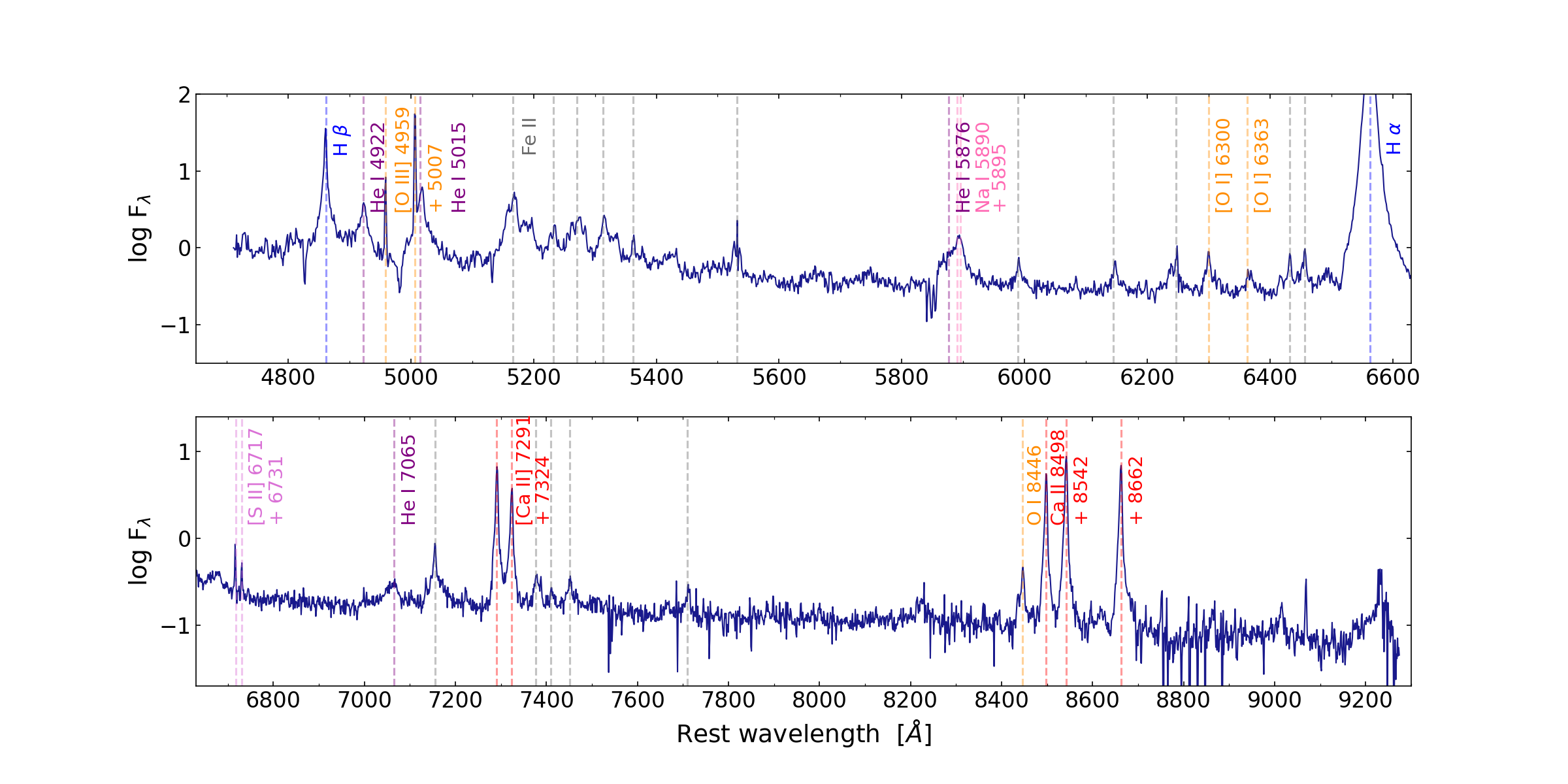}}
\caption{The restframe VLT/MUSE spectrum of SN 2011fh obtained on 2015 May 17, 1359 days after peak brightness. Lines of H, He I, Ca II, Na I, {[O I], [O III], [S II]}, and Fe II are highlighted. Forbidden lines of [O III] and [S II] are most likely due to host galaxy contamination. \label{fig:fig_muse}}
\end{figure*}

Integral field spectroscopy of the galaxy NGC 4806 was obtained with MUSE \citep{2014Msngr.157...13B} on the VLT on 2015 May 17, at a phase of $+1359$ days (program ID:095.D-0172(A); \citealt{2018A&A...613A..35K}). We retrieved the reduced, wavelength and flux calibrated datacube from the ESO archive\footnote{ \url{http://archive.eso.org}}. The cube has $0.2\arcsec \times 0.2\arcsec$ spaxels, a seeing of $0.8\arcsec$, and a spectral range of $4750 - 9350$ \AA. We used a $0.8\arcsec$ radius aperture to extract the spectrum of SN 2011fh from the MUSE datacube. The spectrum is shown in Figure \ref{fig:fig_spec} and in Figure \ref{fig:fig_muse}, where we highlight the emission lines observed at very late times.

\section{Results} \label{sec:resul}

\subsection{Host Galaxy} \label{sec:host_gal}

SN 2011fh is located at $\alpha = 12^h56^m14.009^s$ and $\delta = -29^{\circ}29'54.82"$ (J2000.0), in one of the spiral arms of the nearby galaxy NGC 4806. The redshift for NGC 4806, obtained from the NASA-IPAC Extragalactic Database (NED)\footnote{\url{https://ned.ipac.caltech.edu}}, is $z = 0.00803$ \citep{2003A&A...412...57P}. Using the heliocentric velocity for the galaxy of $v = 2408 \pm 4$ km/s \citep{2003A&A...412...57P}, and the Cosmicflows-3 distance-velocity calculator \footnote{ \url{http://edd.ifa.hawaii.edu/NAMcalculator/}}, which has a flow model consistent with $\textrm{H}_0 = 75 \ \textrm{km} \ \textrm{s}^{-1} \ \textrm{Mpc}^{-1}$ \citep{2020AJ....159...67K}, we get a distance of $D = 30.66$ Mpc. Since the model has a zero-point error of $3 \%$ \citep{2020AJ....159...67K}, we obtained an uncertainty in the distance of $ 0.92 \ \textrm{Mpc}$.
The distance modulus is therefore $\mu = 32.43 \pm 0.06$ mag.

We estimate the redshift at the position of SN 2011fh using the MUSE spectrum of the HII region around the SN. We used Gaussian fitting to estimate the central position of the emission lines of H$\alpha$, H$\beta$, [N II] $\lambda \lambda 6548, 6584$ \AA, [S II] $\lambda \lambda 6716, 6731$ \AA, and [O III] $\lambda \lambda 5007, 4959$ \AA, and measured the local redshift of $z = 0.00812$. This is the value we consider throughout the paper. 

The Milky Way (MW) extinction towards NGC 4806 is $A_V = 0.232$ mag \citep{2011ApJ...737..103S} and $E(B-V) = 0.0748$ mag, with $R_V = 3.1$. We estimate the host galaxy extinction using the high resolution MIKE spectrum. The interstellar Na I doublet (Na I D) absorption lines from the galaxy have equivalent widths (EWs) of $0.1231$ \AA, for Na D1 \ and  $0.1966$ \AA, for Na D2. By scaling these values with the EWs of the MW Na I D absorption features (0.1118 \AA \ and  0.1902\AA, for Na D1 and D2, respectively), we obtained $A_V = 0.24$ mag or $E(B-V) = 0.078$ mag for the host galaxy. 
Thus, using the $2 \%$ photometric error in the measurements of \citet{2011ApJ...737..103S}, and assuming a $10\%$ uncertainty in the host extinction, the total estimated V-band extinction for SN 2011fh is $A_V = 0.48 \pm 0.03 \ \textrm{mag}$.   

We also estimated the extinction of the cluster of stars near SN 2011fh. We use a spectrum extracted with a $2-3 \arcsec$ annulus aperture from the MUSE datacube and measured a Balmer decrement of H$\alpha$/H$\beta = 3.34$. Using \cite{2010AJ....139..694L}, this value implies a total $E(B-V) = 0.16 \ \textrm{mag}$ or $A_V = 0.49 \ \textrm{mag}$. Since this result is consistent with the total estimated extinction for SN 2011fh, we decided to use the same value of $A_V = 0.48 \ \textrm{mag}$ for the stars surrounding the SN.

\subsection{Photometric Evolution} \label{sec:phot_evol}

Figure \ref{fig:fig2} shows the complete photometric evolution of SN 2011fh from 2007 to 2017. The SN reached peak $r$-band brightness on 2011 August 24
(JD = 2455799), at $r = 14.94 \pm 0.02$~mag, or M$_r = -17.88 \pm 0.04$~mag\footnote{For the uncertainty in the absolute magnitudes, we took into account the estimated error in the apparent magnitude, the $3\%$ uncertainty in the distance given by \citet{2020AJ....159...67K}, and the $2\%$ error in color estimated by \citet{2011ApJ...737..103S}.}. Before peak, a five month-long (173 days) brightening event is observed, where the progenitor goes from $r = 17.63 \pm 0.15$ on 2011 February 11, to $r = 16.26 \pm 0.04$ on 2011 July 28. 
After this slow brightening, the source shows a rapid increase to $r \sim 14.94$ mag over $\sim 40$ days, culminating in the August peak. 

\begin{figure}[t!]
\centerline{\includegraphics[scale=.42]{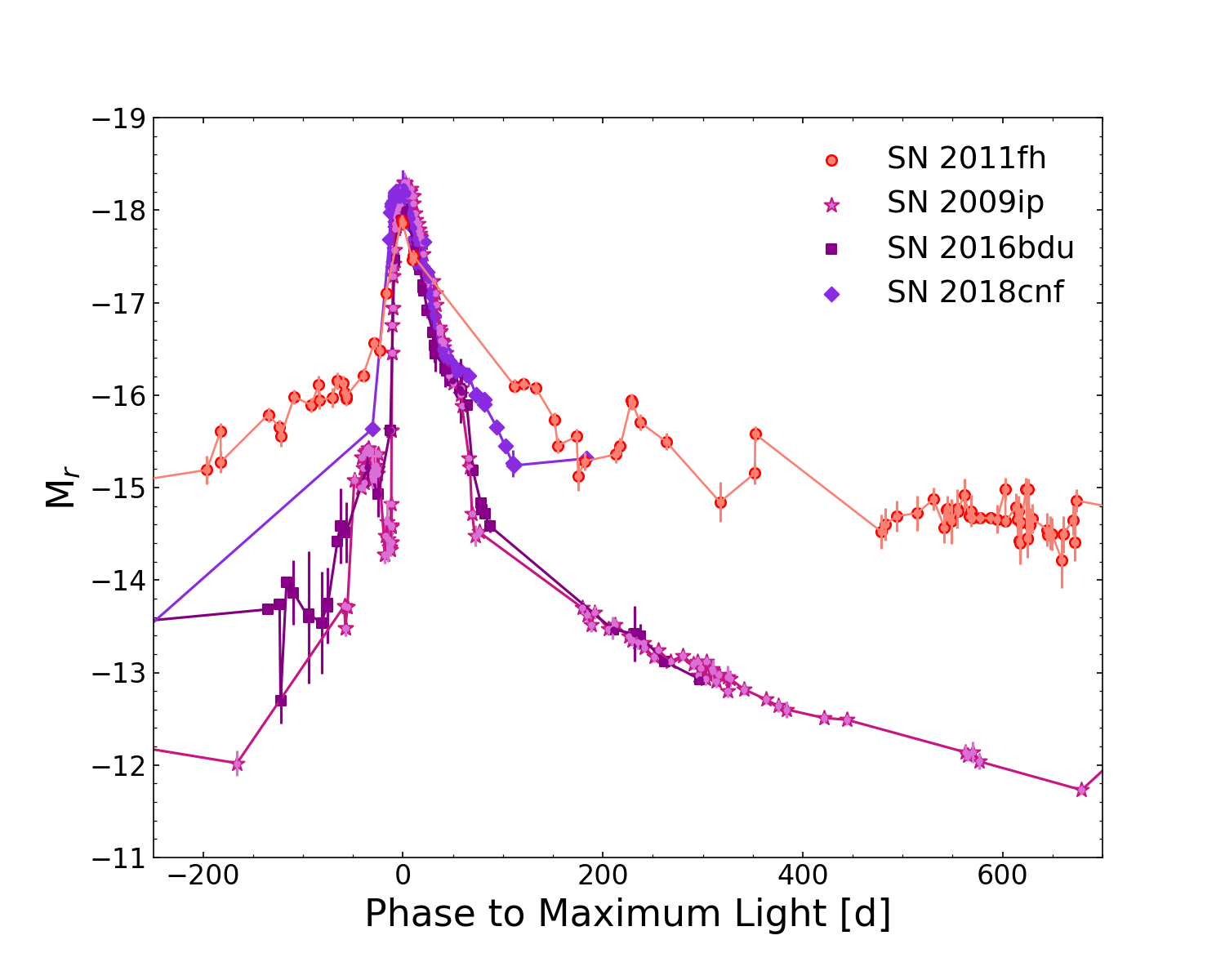}}
\caption{Comparison of the $r$-band light curve near peak brightness of SN 2011fh to SN 2009ip \citep{2013MNRAS.430.1801M, 10.1093/mnras/stt813}, SN 2016bdu \citep{10.1093/mnras/stx2668} and SN 2018cnf \citep{2019A&A...628A..93P}. The phase is in days relative to peak brightness. \label{fig:fig_compare_lc}}
\end{figure}

A precursor event followed by a rapid rise to peak brightness is commonly seen in 2009ip-like transients. For this class of transients, the initial rise is commonly called ``Event A'', and the main peak is called ``Event B''. In Figure \ref{fig:fig_compare_lc}, we show the $r$-band light curve near peak brightness of SN 2011fh, SN 2009ip \citep{2013MNRAS.430.1801M, 10.1093/mnras/stt813}, SN 2016bdu \citep{10.1093/mnras/stx2668}, and SN 2018cnf \citep{2019A&A...628A..93P}.
The Vega $R$-band magnitudes of SN 2009ip were scaled to AB magnitudes by adding $0.21$ mag \citep[][]{2007AJ....133..734B}. 
The four transients show a remarkably similar absolute magnitude at peak, with  M$_r = -18.29 \pm 0.10$ for SN 2009ip, M$_r = -18.00 \pm 0.16$ for SN 2016bdu, and M$_r = -18.25 \pm 0.05$ for SN 2018cnf.
SN 2009ip and SN 2016bdu have a very similar Event A, peaking at M$_r \sim -15.5$, and lasting for $\sim 40$ days. Although different in brightness and duration, the slow increase in luminosity seen for SN 2011fh before peak might share a similar origin to the other transients. After the precursor Event A, all four transients show a rapid brightening to peak, over $\sim 15$ days for SN 2009ip, $\sim 25$ days for SN 2016bdu, and $\sim 30$ days for SN 2018cnf. It is worth noting that the magnitude difference between the end of Event A and the peak brightness is considerably larger for the other transients. 

Before the explosive events of 2011, the source of SN 2011fh is seen to be bright, with an apparently constant luminosity over the preceding three years (see Figure \ref{fig:fig2}). The images are relatively shallow, so it is hard to tell whether the variability is real or not. The absolute magnitudes of M$_r \sim -14.5$ to M$_r \sim -15.5$ are very high in the years preceding the major brightening event. For comparison, the progenitors of SN 2009ip and SN 2016bdu had $r$-band absolute magnitudes of $-13$ to $-14$ mag. 

The stellar cluster around SN 2011fh might be a source of photometric contamination in the pre-outburst images, which were all obtained using cameras with large pixel scales. 
We estimate the photometric contamination in Appendix \ref{sec:contam}. Our analysis shows that the contribution of the stellar cluster is not a serious problem, although it increases for the bluer bands. 

Spitzer imaging from 2009 also reveals a bright source at the position of SN 2011fh. Photometry of the 3.6- and 4.5-micron channels show a source with apparent magnitudes of 15.9 and 15.5, respectively. These imply luminosities of $3.5 \times 10^6$ and $2.5 \times 10^6$ L$_\odot$, respectively. In the discovery report, \cite{2011CBET.2799....1M} noted that the mid-IR luminosity may have a contribution from the HII region around the source, due to the 260 pc spatial resolution of the IRAC instrument. 

During the first days after peak, the brightness of SN 2011fh drops significantly. Unfortunately, between 7 and 110 days after peak in $r$-band, there were no photometric observations of the source. After 110 days, it is significantly fainter and slowly fading. The decay, however, is not monotonic, with two pronounced peaks at 228 (2012 April 11, or JD = 2456029) and 350 days (2012 August 12, or JD = 2456151) after peak (Figure \ref{fig:fig_compare_lc}). Bumps and fluctuations after peak brightness are common in the light curve of SN 2009ip-like objects, but the amplitudes seen for the other three sources shown in Figure \ref{fig:fig_compare_lc} are significantly smaller. This behavior is usually interpreted as the collision of the ejecta with circumstellar material showing a complex density profile. 
The overall evolution of SN 2011fh in the months following peak brightness is, however, very similar to SN 2009ip, with a slow fading probably caused by the ongoing interaction with a dense CSM.

The CSP2 optical $BVri$- and NIR $YJH$-band images of SN 2011fh, as well as new IR Spitzer observations, were obtained only starting $\sim 550$ days after peak (2013 February 20). The behavior in the other filters is very similar to $r$-band. Initially, all bands show a slow drop in brightness, with an apparent flattening by the beginning of 2015, 1300 days after peaking in $r$-band. At this epoch the source is still quite luminous, with M$_r \sim -13.5$ mag. Surprisingly, an apparent rise in luminosity is detected by the beginning of 2016 in the optical $BVi$- and NIR $YJH$-bands, although not in the $r$-band. The apparent brightening in the NIR bands is clear in the last point of the light curve where the $Y$-band brightened by $\Delta Y \approx 0.36$ mag over 418 days.

HST observations from October 2016 show a very bright source at the position of SN 2011fh (Figure \ref{fig:fig_hst}), with absolute magnitudes of M$_{F814W} = -13.32 \pm 0.05$ mag and M$_{F336W} = -14.25 \pm 0.09$ mag. This is much more luminous than even the most luminous quiescent stars known \citep[e.g., R136a1 and BAT99-98 in the Tarantula Nebula of the Large Magellanic Cloud, with absolute bolometric magnitudes of $-12.2$ and $-12.0$, respectively;][]{2020MNRAS.499.1918B, 2014A&A...565A..27H}. Another parallel can be made with Eta Carinae, which reached an absolute magnitude M$_V \sim -14$ mag during its 19th Century Great Eruption \citep{2018MNRAS.480.1466S}. 
This might be simply due to late-time emission due to the ongoing interaction of the ejecta with the CSM, as it is commonly seen for interacting transients \citep[e.g.,][]{2017MNRAS.469.1559G}, or alternatively indicate that we are seeing the star in a LBV-like eruptive state.

\subsubsection{Colors} \label{sec:colors}

\begin{figure}[t!]
\centerline{\includegraphics[scale=.5]{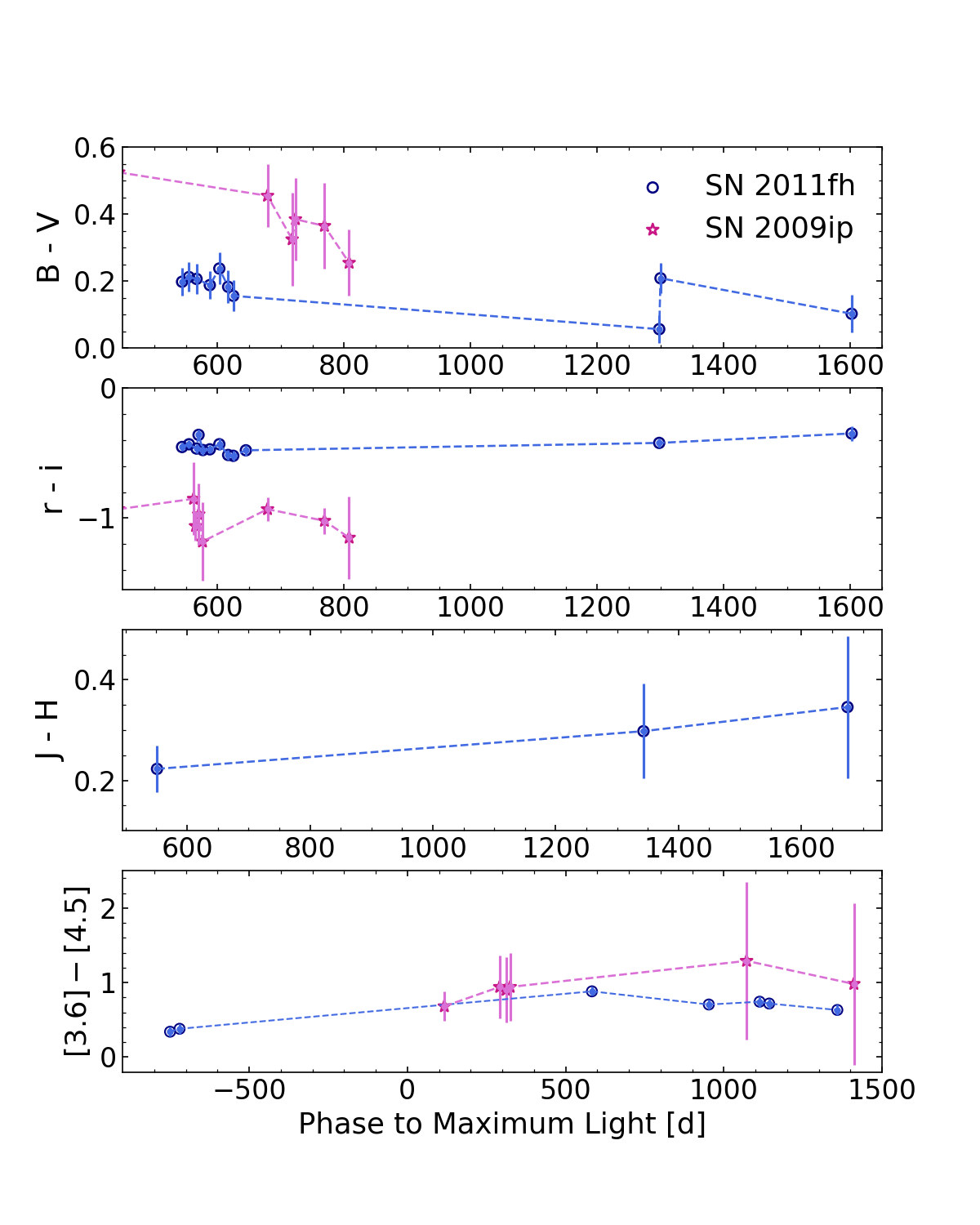}}
\caption{$B-V$, $r-i$, $J-H$ and $[3.6]  - [4.5]$ color evolution of SN 2011fh. Phase is in days relative to $r$-band peak. For comparison, we also show the $B-V$, $R-I$, and $[3.6] - [4.5] $ color evolution of SN 2009ip \citep{2015MNRAS.453.3886F, 2019ApJS..241...38S}. We show the $R-I$ color, because $r-i$ data is not available. \label{fig:fig_color_evol}}
\end{figure}

Figure \ref{fig:fig_color_evol} shows the CSP2 $B-V$, $r-i$, $J-H$, and the $[3.6] - [4.5]$ color evolution of SN 2011fh. For the optical and near-IR bands, unfortunately, we only have colors at very late-times, starting at 550 days after the maximum brightness. We are able to estimate the mid-IR colors of the progenitor 750 days before peak brightness.

As commonly seen for Type IIn SNe, the $B-V$ and $r-i$ evolution of SN 2011fh is very slow \citep{2013A&A...555A..10T}. The $B-V$ color becomes slightly bluer, consistent with the temperature increase measured in the spectra and the overall spectral energy distribution (SED, see Section \ref{sec:sed} and Section \ref{sec:spec_evol}). The $J-H$ color becomes slightly redder, increasing from 0.22~mag, at phase $+550$ days, to 0.34~mag, at phase $+1676$ days, although this last value has large uncertainties. 

Type IIn SNe are usually bright in the mid-IR bands, and have very heterogeneous properties \citep[][]{2021arXiv210612427S}. 
A significant color evolution is seen for SN 2011fh in the mid-IR, if we compare the values before and after the luminous outburst of 2011. The source has $[3.6]  - [4.5]  = 0.37$~mag at phase $-715$ days, and $[3.6]  - [4.5]  = 0.87$~mag, 580 days after peak. After peak, the mid-IR color becomes a little redder, decreasing slowly with time. 

We also show the $B-V$ and $R-I$ evolution of SN 2009ip in Figure \ref{fig:fig_color_evol} from \cite{2015MNRAS.453.3886F}. We show the $R-I$ colors, because $r-i$ is not available for these epochs. 
The Vega R- and I-band values were scaled to AB magnitudes by adding, respectively, $0.21$~mag and $0.45$~mag \citep[][]{2007AJ....133..734B}.
In $B-V$, SN 2011fh is significantly bluer than SN 2009ip, although both events become redder with time. This might be due to the fact that SN 2011fh is located in a blue star-forming region, with the contamination from the cluster increasing as the SN fades (see Appendix \ref{sec:contam}). The other possible explanation is that SN 2011fh is intrinsically hotter than SN 2009ip. The $R-I$ colors of SN 2009ip are redder than the $r-i$ values for SN 2011fh. This might imply that SN 2009ip has stronger H$\alpha$ emission lines in relation to the spectral continua throughout its evolution. SN 2009ip has a very similar mid-IR color evolution to SN 2011fh. Although the mid-IR estimates are uncertain at late-times, SN 2009ip becomes a little redder after peak, and then bluer around a phase of $\sim 1000$ days \citep[][]{2019ApJS..241...38S, 2021arXiv210612427S}.

\subsubsection{Spectral Energy Distribution} \label{sec:sed}

\begin{figure}[t!]
\centerline{\includegraphics[scale=.5]{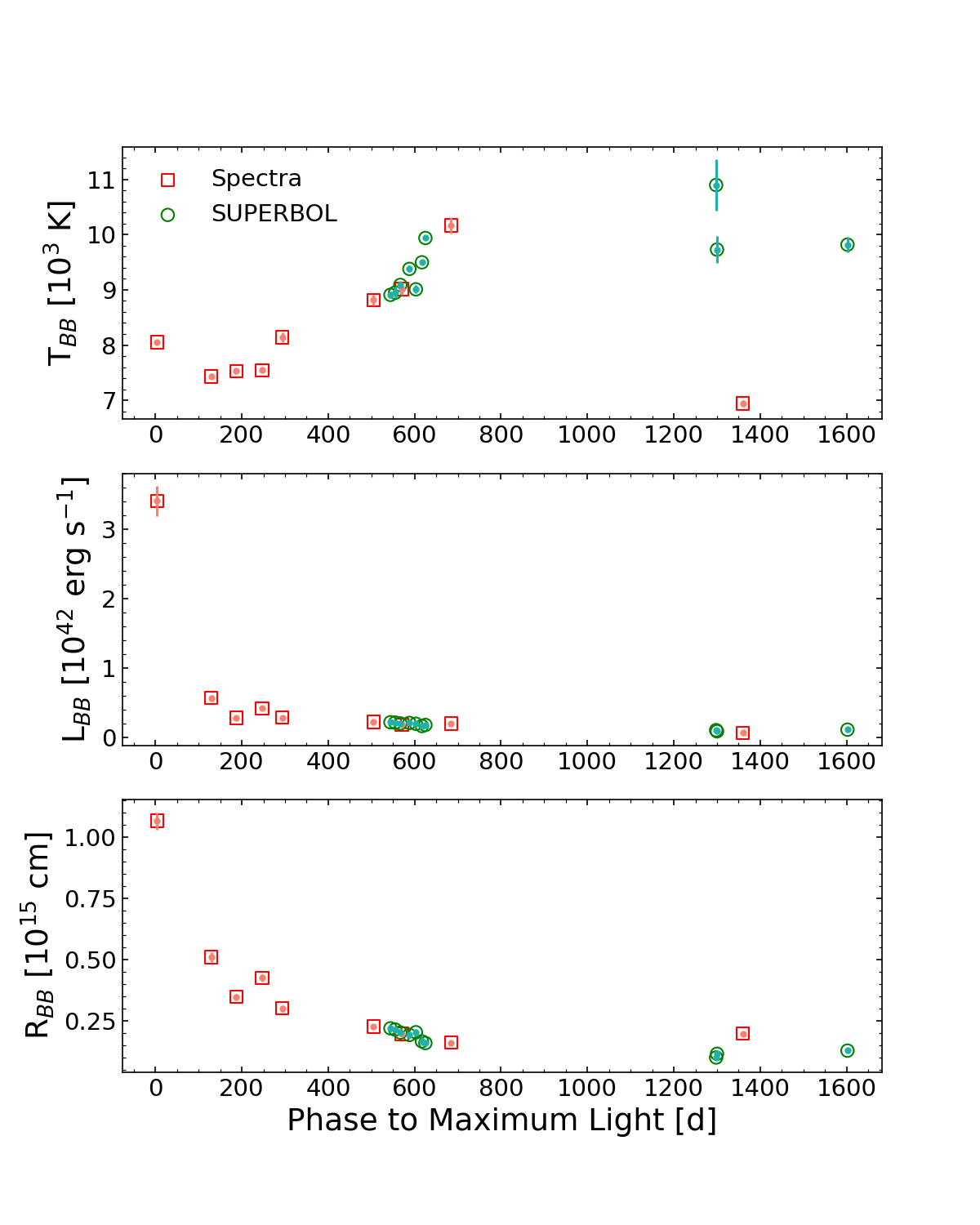}}
\caption{Temperature (T$_{BB}$), bolometric luminosity (L$_{bol}$), and radius (R$_{BB}$) evolution of SN 2011fh, estimated from blackbody fits to the spectral continua using {\tt\string Astropy} models (red squares), and to the spectral energy distributions (SEDs) using {\tt\string SUPERBOL} (green circles). The phase is relative to the maximum $r$-band luminosity in days.\label{fig:fig_bb_superbol}}
\end{figure}

We use the optical, near-infrared (NIR), and infrared (IR) photometry to constrain the spectral energy distribution (SED) of SN 2011fh after peak. We used {\tt\string SUPERBOL} \citep[Version 1.7;][]{2018RNAAS...2..230N} to estimate the blackbody temperature, luminosity, and photosphere radius between phases $+544$ and $+1602$ days. Since SN 2011fh has very strong H$\alpha$ emission (see Section \ref{sec:spec_evol}), we carried out this first analysis using only the optical $BVi$ bands. The results of the single-blackbody fits to these three filters are shown in Figure \ref{fig:fig_bb_superbol} and reported in Table \ref{tab:SUPERBOL}.

SN 2011fh has an unusual temperature evolution for either a SN or a SN impostor. Transients driven by CSM interaction usually show a smooth decrease in temperature after peak \citep{2014ApJ...780...21M, 2018MNRAS.473.4805K}, opposite to the behavior of SN 2011fh. The blackbody temperature at phase $+544$ days is T$_{BB} \approx 8920$~K, and it rises to T$_{BB} \approx 9950$~K at phase $+630$ days. A higher blackbody temperature of $\sim 11000$~K is measured at 1300 days, but the late-time measurements are uncertain, and the fluxes may be dominated by line emission at these epochs (see Figure \ref{fig:fig_spec} and Section \ref{sec:spec_evol}). The radius evolution of SN 2011fh also differs from the usual increase and then smooth decrease after peak. The evolution of the photospheric radius of SN 2009ip was also unusual and was interpreted by \cite{2014ApJ...780...21M} as being due to a more complex CSM structure.

\begin{figure}[h!]
\centerline{\includegraphics[scale=.40]{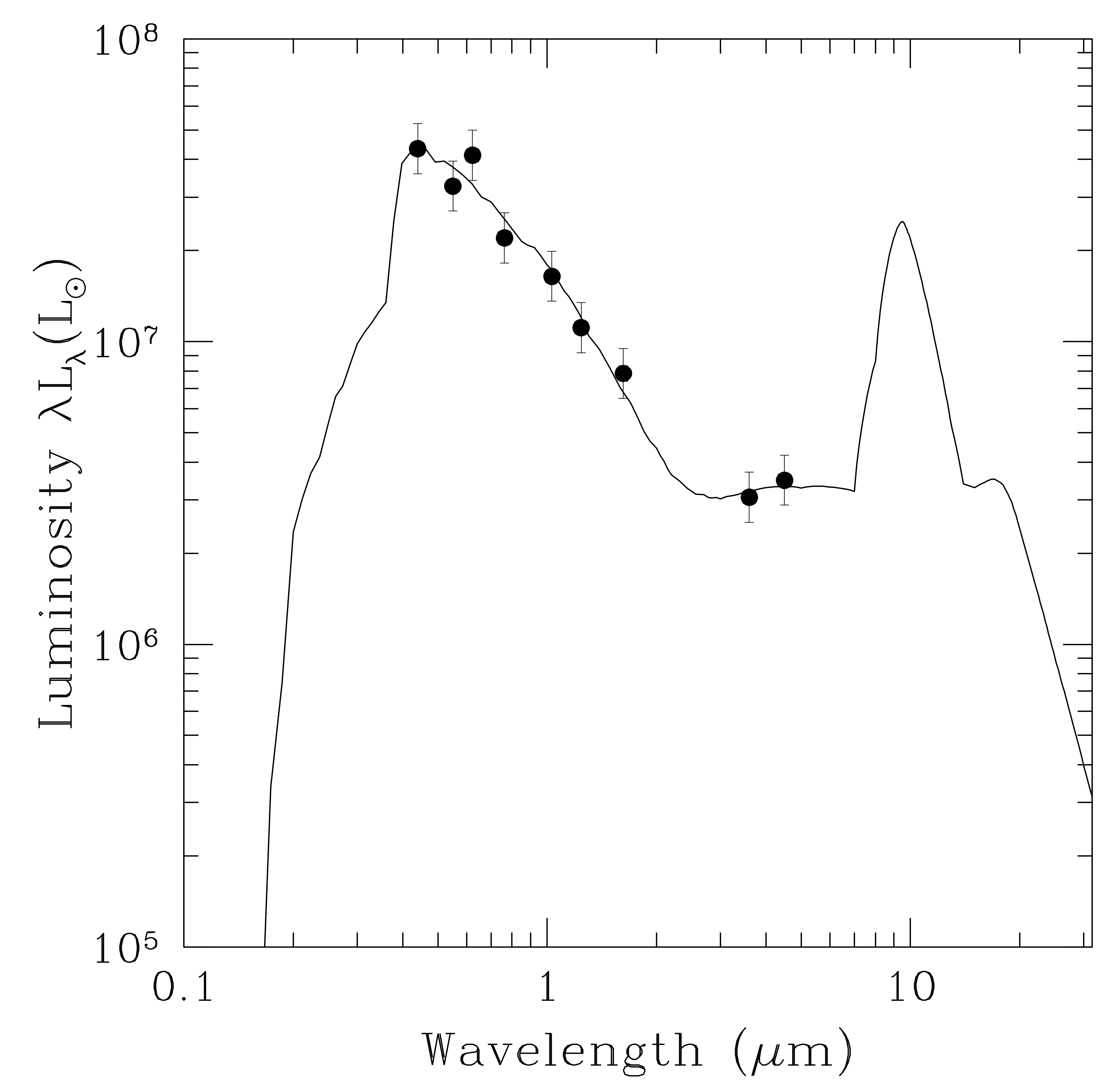}}
\caption{{\tt\string DUSTY} fit to the SED at phase $+570$, using the optical and Spitzer data. The photospheric temperature is $\sim 7865$~K and the dust temperature is $\sim 950$~K. The fit parameters are given in Table \ref{tab:dusty_fits}. \label{fig:dusty_fit_570}}
\end{figure}

We also model the SED using {\tt\string DUSTY} \citep{1997MNRAS.287..799I}, a
code for solving radiative transfer through a spherically symmetric
dusty medium. We assume a central stellar source using atmospheric models from \citet{2003IAUS..210P.A20C}, a dust shell with a thickness $R_{out}/R_{in} = 2$, and silicate dust from \citet{1984ApJ...285...89D} with a standard MRN dust size distribution \citep{1977ApJ...217..425M}. We find best-fitting models using a MCMC wrapper around  {\tt\string DUSTY} as in \citet{2015MNRAS.452.2195A}.  In Table \ref{tab:dusty_fits} we show the resulting photospheric ($T_{phot}$) and dust ($T_{dust}$) temperatures, together with the estimated photospheric radius ($R_{phot}$) and inner radius of the dust shell ($R_{in, dust}$), the total luminosity ($L$), and the dust optical depth in V band ($\tau_V$). 
Figure \ref{fig:dusty_fit_570} shows the optical, NIR and IR SED, and the {\tt\string DUSTY} fits at phase $+570$ days.
At this period, the SED was fit with $\tau_V = 0.68$, $T_{phot} \approx 7865$~K, $T_{dust} \approx 950$~K, and $L \approx 10^{7.74} \  \textrm{L}_\odot$.
The dust emission increases in temperature with time, reaching  $\sim 1116$~K at phase $+1143$ days.

\begin{deluxetable*}{cccccccccccc}
\tablecaption{DUSTY fits of the SEDs \label{tab:dusty_fits}}
\tablewidth{0pt}
\tablehead{
\colhead{Phase} & \colhead{$\tau_V$} & \multicolumn2c{$T_{phot}$} & \multicolumn2c{$T_{dust}$}  & \multicolumn2c{log $L$} & \multicolumn2c{log $R_{phot}$} & \multicolumn2c{log $R_{in, dust}$}\\
\colhead{(days)} & \colhead{}  & \colhead{(K)} & \colhead{($1\sigma$ error)} & \colhead{(K)} & \colhead{($1\sigma$ error)} & \colhead{(L$_\odot$)} & \colhead{($1\sigma$ error)} & \colhead{(R$_\odot$)} & \colhead{($1\sigma$ error)} & \colhead{(cm)} & \colhead{($1\sigma$ error)} 
}
\decimalcolnumbers
\startdata
+570 &  0.68 & 7865 & 511 & 950 & 320 & 7.74 & 0.04 & 3.60 & 0.04 & 16.52 & 0.29\\ 
+1113 & 0.91 & 8321 & 537 & 903 & 233 & 7.48 & 0.06 & 3.42 & 0.03 & 16.50 & 0.26\\ 
+1143 & 0.61 & 8823 & 779 & 1116 & 240 & 7.48 & 0.05 & 3.37 & 0.05 & 16.38 & 0.18\\
+1300 & 0.78 & 8915 & 635 & 1006 & 182 & 7.40 & 0.05 & 3.32 & 0.04 & 16.44 & 0.17\\
\enddata

\end{deluxetable*}

\begin{deluxetable}{rcccccc}
\tablecaption{SUPERBOL fits to the $BVi$-bands \label{tab:SUPERBOL}}
\tablewidth{0pt}
\tablehead{
\colhead{Phase} & \multicolumn2c{T$_{BB}$} & \multicolumn2c{$10^{41}$ L$_{BB}$} & \multicolumn2c{$10^{14}$ R$_{BB}$}\\
\colhead{(days)} & \colhead{(K)} & \colhead{(error)} & \colhead{(erg s$^{-1}$ )} & \colhead{(error)} & \colhead{(cm)} & \colhead{(error)}
}
\decimalcolnumbers
\startdata
+544 & 8910 & 55 & 2.17 & 0.11 & 2.20 & 0.15 \\
+555 & 8950 & 105 & 2.12 & 0.15 & 2.15 & 0.15 \\
+566 & 9090 & 74 & 2.00 & 0.10 & 2.03 & 0.14 \\
+587 & 9380 & 58 & 2.07 & 0.27 & 1.93 & 0.16 \\
+602 & 9010 & 76 & 1.95 & 0.24 & 2.04 & 0.14 \\
+616 & 9500 & 38 & 1.62 & 0.17 & 1.67 & 0.16 \\
+625 & 9940 & 27 & 1.78 & 0.08 & 1.60 & 0.11 \\
+1298 & 10900 & 465 & 1.03 & 0.11 & 1.01 & 0.08 \\
+1300 & 9730 & 246 & 0.86 & 0.05 & 1.16 & 0.08 \\
+1602 & 9820 & 147 & 1.10 & 0.05 & 1.29 & 0.07 \\
\enddata

\end{deluxetable}

\subsection{Spectroscopic Evolution} \label{sec:spec_evol}

\begin{figure}[t!]
\centerline{\includegraphics[scale=.4]{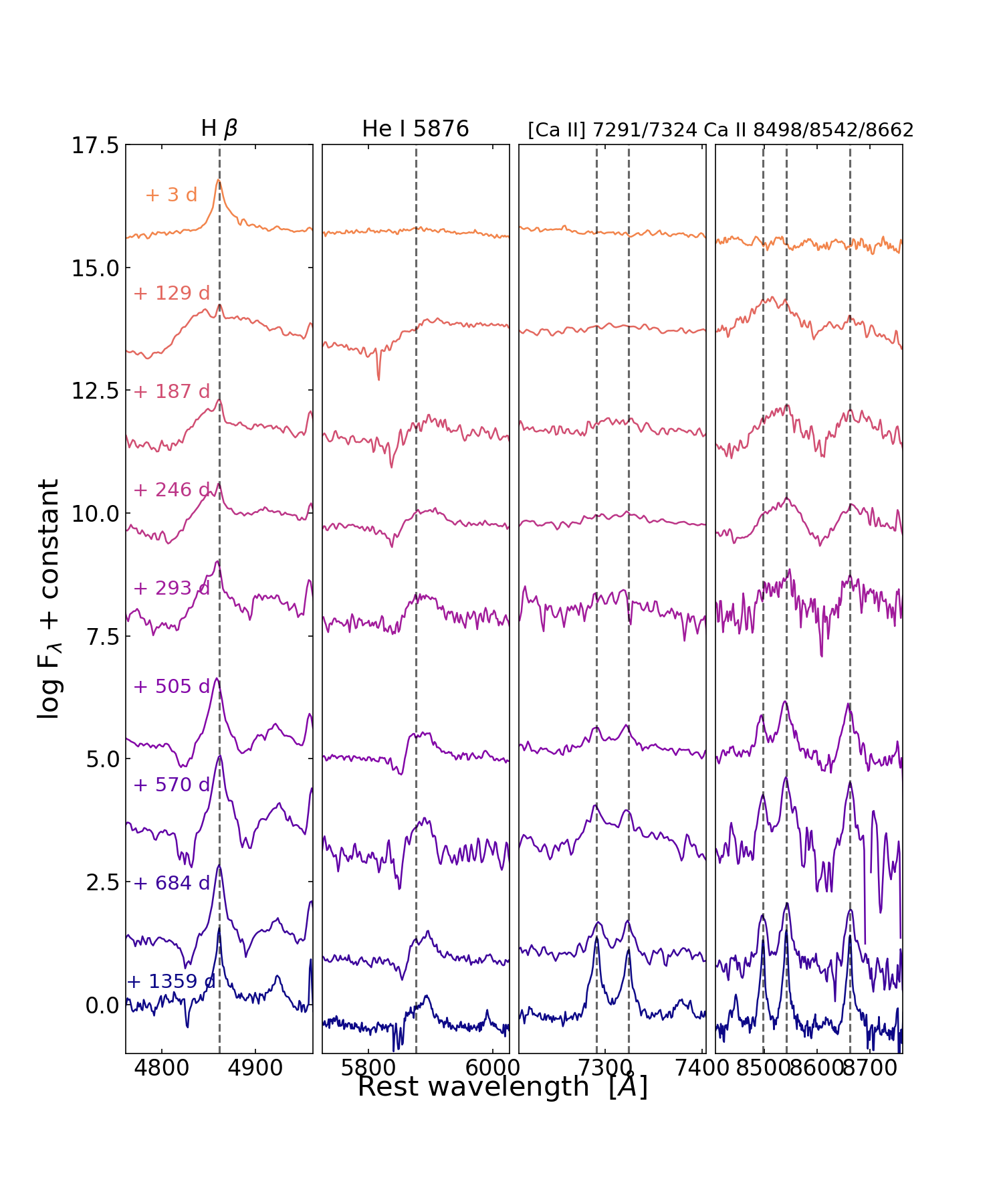}}
\caption{Evolution of H$\beta$, He I $\lambda 5876$, the [Ca II] $\lambda \lambda 7291, 7324$ doublet, and the Ca II $\lambda \lambda 8498, 8542, 8662$ triplet. The vertical dashed lines mark the rest wavelengths.  \label{fig:fig_HB_He_Ca}}
\end{figure}

In Figure \ref{fig:fig_spec}, we show the complete spectral evolution of SN 2011fh from 3 to 1359 days after peak brightness. We use vertical lines to highlight prominent emission features of H, He I, Ca II, Na I, [S II], and [O II]. The spectra are dominated by Balmer lines in emission during the whole evolution, with features of He I and Ca II becoming more prominent with time. The forbidden lines of [S II] and [O II] seen throughout the evolution of SN 2011fh are most likely due to host galaxy contamination. In Figure \ref{fig:fig_HB_He_Ca}, we show the evolution of the emission profiles of H$\beta$, He I $\lambda 5876$, the [Ca II] $\lambda \lambda 7291, 7324$ doublet, and the Ca II $\lambda \lambda 8498, 8542, 8662$ triplet. 

\begin{figure}[t!]
\centerline{\includegraphics[scale=.35]{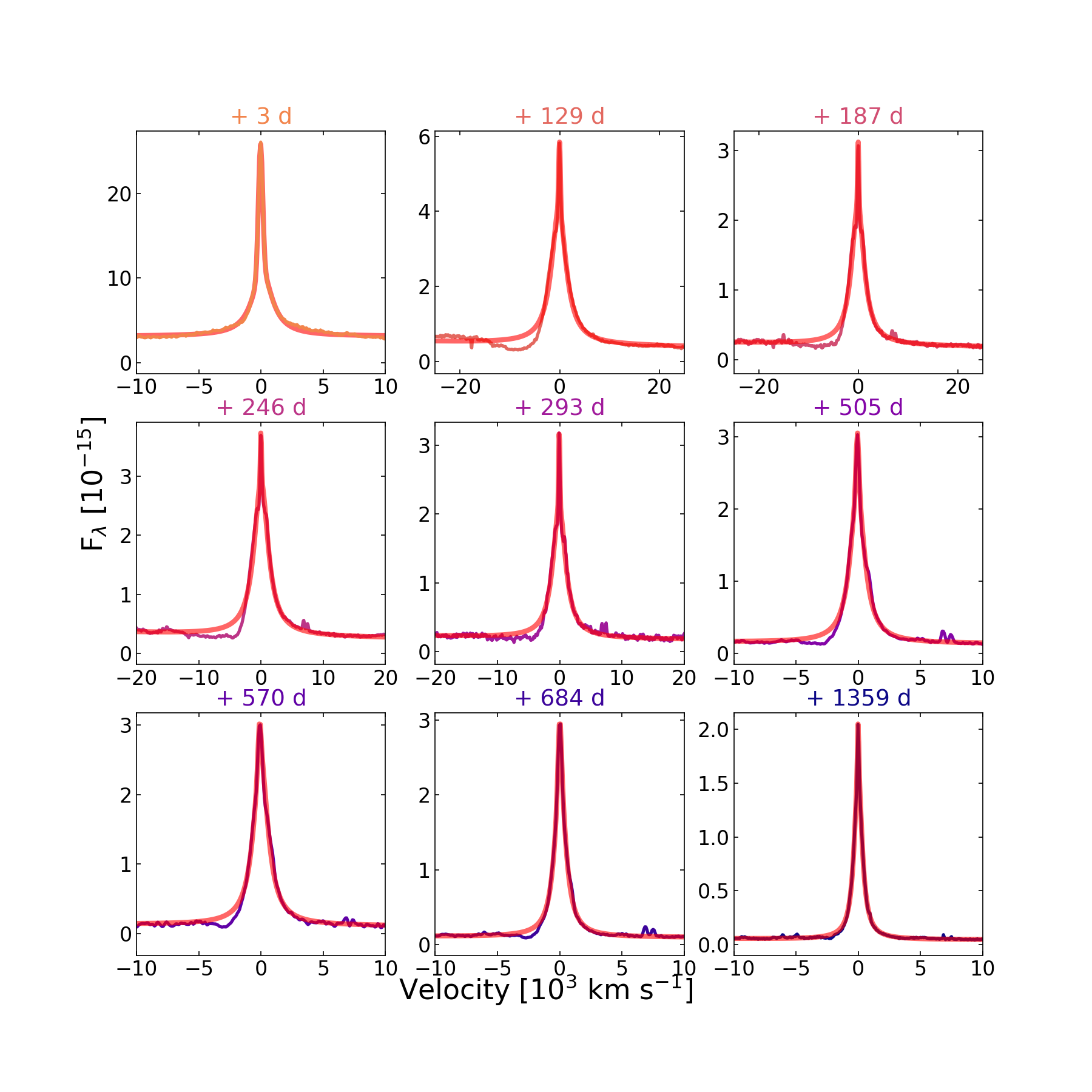}}
\caption{Evolution of the H$\alpha$ emission line and a model built with a narrow-width Gaussian component and an intermediate-width Lorentzian component. Velocity corresponds to rest-frame wavelength of H$\alpha$, and the spectral phases are given above each panel. Absorption features are seen in some epochs, especially at $+ 129$ and $+ 187$ days. \label{fig:fig_Ha_fit}}
\end{figure}

\begin{figure}[t!]
\centerline{\includegraphics[scale=.5]{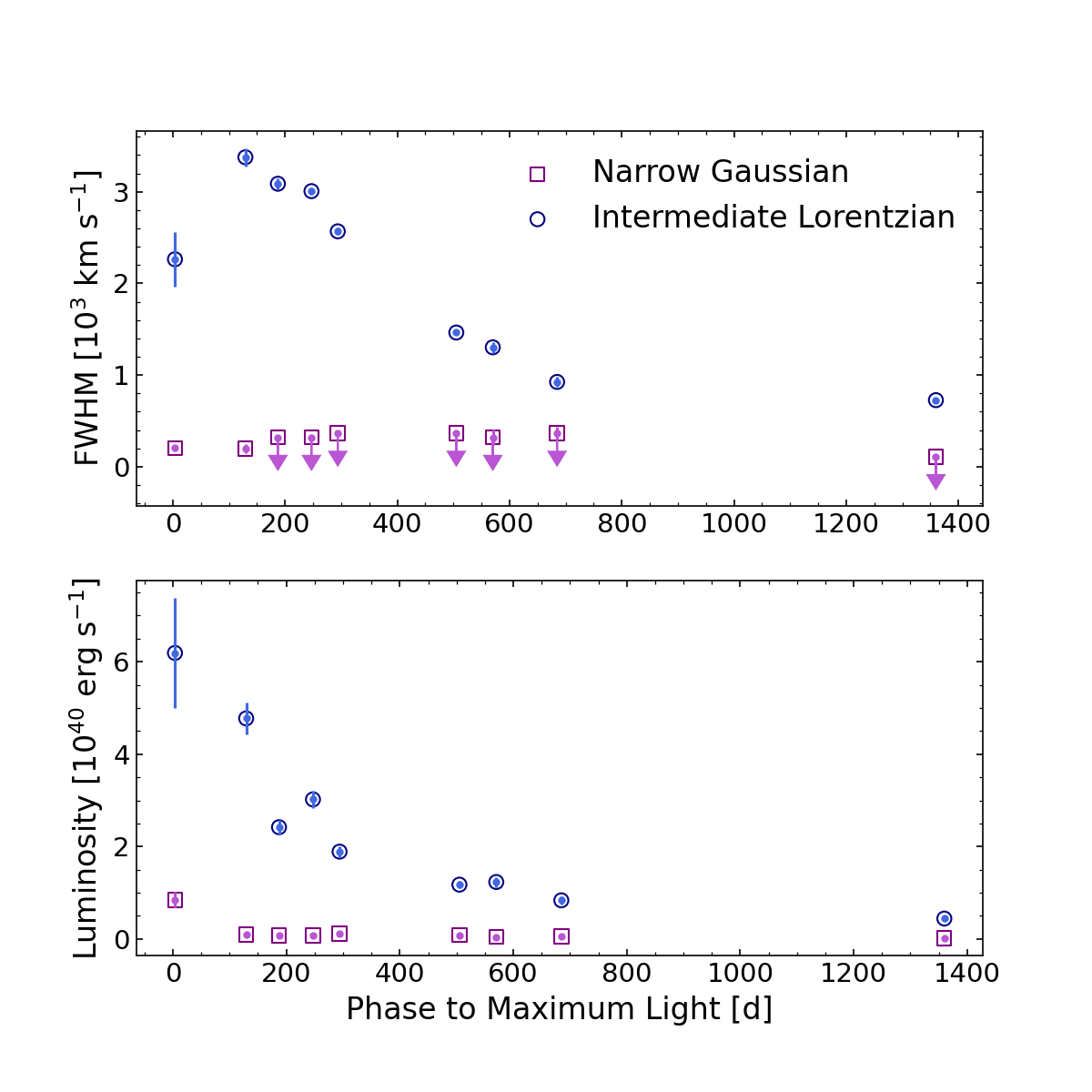}}
\caption{Evolution of the H$\alpha$ FWHM velocity (top panel) and luminosity (bottom panel). The purple squares are the narrow Gaussian component, while the blue circles are the Lorentzian. The arrows indicate an upper limit for the narrow component, given by the spectral resolution. Phase is in days relative to $r$-band maximum. \label{fig:fig_Ha_fwhm_L}}
\end{figure}

Due to the different emission mechanisms, interacting transients can show very complex Balmer line profiles \citep[see the review by][]{2017hsn..book..403S}. For the spectra of SN 2011fh, we used {\tt\string Astropy} \citep{astropy:2013, astropy:2018} to fit the H$\alpha$ emission profiles with narrow Gaussian components and intermediate Lorentzian components, plus a constant continuum, as shown in Figure \ref{fig:fig_Ha_fit}. 
The estimated full width at half maximum (FWHM) velocity and the luminosity of the two components throughout the evolution of SN 2011fh are shown in Figure \ref{fig:fig_Ha_fwhm_L}, and reported in Table \ref{tab:Ha_FWHM} and Table \ref{tab:Ha_lum}. Errors are estimated by the $1\sigma$ parameter uncertainty for the best-fitting Lorentzian and Gaussian models. A $3\%$ uncertainty is also considered for the error in the luminosity of the components. Due to its proximity to the spectral resolution, the FWHM of the narrow components are corrected by the resolution value. At some epochs, the width of the narrow component falls below that value, and are indicated as less than the spectral resolution in Table \ref{tab:Ha_FWHM} and Figure \ref{fig:fig_Ha_fwhm_L}. Prominent blue absorption features are also seen at certain epochs, especially at phases +129 and +187 days. The photospheric temperature, radius and bolometric luminosity evolution of SN 2011fh, obtained from the spectral continuum fitting, also shown in Figure \ref{fig:fig_bb_superbol} and reported in Table \ref{tab:BB_fits}, agree well with the SED models. Errors in temperature are given by the $1\sigma$ parameter uncertainty for the best-fitting model, while for the error in luminosity and radius, a $3\%$ uncertainty in the distance is also considered.

The first spectrum in our sequence, obtained 3 days after peak, is dominated by Balmer lines in emission, and resembles a typical Type IIn SN. It has a blue continuum with a blackbody temperature of $ T_{BB} \sim 8000 \ \textrm{K}$. The H$\alpha$ profile can be fit with an intermediate-width Lorentzian profile with a FWHM velocity of $v_{FWHM} \sim 2263 \ \textrm{km s}^{-1}$, and a narrow Gaussian component with  $v_{FWHM} \sim 204 \ \textrm{km s}^{-1}$.
The narrow Gaussian emission is generated by the ionized dense CSM, while the Lorentzian profile is indicative of broad electron scattering in the unshocked CSM \citep{2001MNRAS.326.1448C}.

The H$\alpha$ profile becomes significantly broader in the second spectrum at $+ 129$ days. The Lorentzian profile now has $v_{FWHM} \sim 3376 \ \textrm{km s}^{-1}$, and the narrow Gaussian has $v_{FWHM} \sim 197 \ \textrm{km s}^{-1}$. The line is well described by a P-Cygni profile with a strong blueshifted absorption feature. The blueshifted absorption wing extends to $\sim 6430 \ \textrm{km s}^{-1}$, and is probably formed by fast-moving material along the line of sight. At this epoch, the He I features and the Ca II triplet start to develop, with H$\beta$ blended into the He I $\lambda 4922$ emission line. The continuum becomes relatively redder, with a blackbody temperature of $ T_{BB} \sim 7400 \ \textrm{K}$. 

The FWHM of the H$\alpha$ Lorentzian begins to decrease at $+ 187$ days, with $v_{FWHM} \sim 3087 \ \textrm{km s}^{-1}$, and the narrow Gaussian component is below the spectral resolution of $\sim 320 \ \textrm{km s}^{-1}$. The blueshifted absorption feature is still present, with a velocity of $\sim 6447\ \textrm{km s}^{-1}$. The blackbody temperature remains almost constant, with $ T_{BB} \sim 7500 \ \textrm{K}$. 

By $293$ days, the P-Cygni absorption feature is no longer easy to identify, although the H$\alpha$ line cannot be entirely described by only emission features, as can be seen in Figure \ref{fig:fig_Ha_fit}. We measure $v_{FWHM} \sim 2568 \ \textrm{km s}^{-1}$ for the Lorentzian component. This epoch is marked by a sudden inversion in temperature to $ T_{BB} \sim 8100 \ \textrm{K}$. At this epoch, the emission features of Ca II become more prominent, with the [Ca II] $\lambda \lambda 7291, 7324$ doublet beginning to develop and the three components of the Ca II $\lambda \lambda 8498, 8542, 8662$ triplet becoming easily distinguishable. 

The emission features remain very similar during the three next epochs, with He I and Ca II now very prominent. At phase $+505$ days, the Ca II $\lambda8662$ line can be fit with a narrow Lorentzian profile with $v_{FWHM} \sim 583 \ \textrm{km s}^{-1}$. The Fe II emission line forest is also detected, and becomes stronger in the last spectrum of our sequence (see Figure \ref{fig:fig_muse}). At $+ 684$ days, the H$\alpha$ profile gets considerably narrower, with $v_{FWHM} \sim 925 \ \textrm{km s}^{-1}$ for the Lorentzian component. At this time, the blackbody temperature reaches its maximum measured value of $ T_{BB} \sim 10200 \ \textrm{K}$.

The unusual temperature increase seen in SN 2011fh, to both SEDs and the spectral continua, might be generated by the shock interacting with the CSM, a scenario that might be supported by the bumps seen in the $r$-band light curve. Another possibility is that the CSM is becoming relatively thin, revealing a hot stellar photosphere.

The NIR spectrum taken on 2013 April 18, 600 days after peak (Figure \ref{fig:spec_fire}) shows prominent emission lines of the Paschen and Brackett series, and features from He I and Ca II. This was also observed in early NIR spectroscopy of the Event B in SN 2009ip \citep{2013ApJ...767....1P, 2013MNRAS.434.2721S, 2014ApJ...780...21M}. Although our spectrum was taken in a much later time after peak, it is still strongly dominated by emission features of hydrogen, such as it was observed in 2012 for SN 2009ip.

The very late-time optical spectrum of SN 2011fh, obtained on 2015 October 7 at phase $+1359$ days, is shown in Figure \ref{fig:fig_muse}. We use {\tt\string The Atomic Line List v2.04} \footnote{ \url{https://www.pa.uky.edu/~peter/atomic/}} to identify the emission lines in the spectrum, and highlight the main features with traced lines. The Fe II emission features (gray traced lines) become very prominent especially near 5200 \AA. 
At this point, the H$\alpha$ profile is very narrow, being described by a Lorentzian component with $v_{FWHM} \sim 726 \ \textrm{km s}^{-1}$ and a Gaussian component with $v_{FWHM} <110 \ \textrm{km s}^{-1}$. 
The Ca II emission lines also get considerably narrower, with $v_{FWHM} \sim 240 \ \textrm{km s}^{-1}$ for Ca II $\lambda8662$ and the two components of the [Ca II] $\lambda \lambda 7291, 7324$ doublet. 
The detected [O II], [O III], and [S II] lines are due most likely to a background H II region, and become stronger with time as the contribution from SN 2011fh becomes fainter, while the [O I] lines are probably due to the transient.
As shown in Figure \ref{fig:fig_bb_superbol}, we measure a sudden drop in temperature at this epoch, which is not seen for the SUPERBOL measurements in photometry. This difference might be due to the last spectrum being strongly dominated by emission lines, which makes the blackbody fit in the continuum very uncertain.
We also fit the H$\beta$ emission lines with intermediate- and narrow-witdh Lorentzian profiles, and use the resultant luminosities to calculate the Balmer decrement evolution. We estimate a relatively low value of $\textrm{H}\alpha/\textrm{H}\beta = 3.01$ near peak brightness, which might be due to Case B recombination or a high density CSM, as it was observed for 2009ip \citep{2014AJ....147...23L}. For the phase $+246$ days, however, we measure $\textrm{H}\alpha/\textrm{H}\beta \simeq 13$, which is very similar to the value found for the 2009ip-like SN 2016bdu \citep{10.1093/mnras/stx2668} and other Type IIn SNe \citep[e.g., 1996al;][]{2016MNRAS.456.3296B}. Such high ratios might be related to newly formed dust or to Case C recombination where the CSM is optically thick to the Balmer emission lines.

\begin{deluxetable}{ccccc}
\tablecaption{FWHM velocities of the two H$\alpha$ components \label{tab:Ha_FWHM}}
\tablewidth{0pt}
\tablehead{
\colhead{Phase} & \multicolumn2c{v$_{FWHM,  Lorentzian}$} & \multicolumn2c{v$_{FWHM,  Gaussian}$}\\
\colhead{(days)} & \colhead{(km s$^{-1}$)} & \colhead{($1\sigma$ error)} & \colhead{(km s$^{-1}$)} & \colhead{($1\sigma$ error)}
}
\decimalcolnumbers
\startdata
+3 & 2263.33 & 299.11 & 204.03 & 36.12 \\
+129 & 3376.06 & 99.98 & 197.69 & 45.53 \\
+187 & 3087.12 & 60.19 & $<$320.00 & 26.19 \\
+246 & 3006.35 & 42.37 & $<$320.00 & 22.12 \\
+293 & 2568.02 & 40.41 & $<$365.00 & 13.23 \\
+505 & 1464.20 & 23.71 & $<$365.00 & 14.67 \\
+570 & 1302.74 & 59.22 & $<$320.00 & 83.59 \\
+684 & 925.26 & 56.35 & $<$365.00 & 68.59 \\
+1359 & 726.56 & 14.63 & $<$110.00 & 12.78 \\
\enddata

\end{deluxetable}

\begin{deluxetable}{ccccc}
\tablecaption{Luminosities of the two H$\alpha$ components \label{tab:Ha_lum}}
\tablewidth{0pt}
\tablehead{
\colhead{Phase} & \multicolumn2c{$10^{39}$ L$_{Lorentzian}$} & \multicolumn2c{$10^{39}$ L$_{Gaussian}$}\\
\colhead{(days)} & \colhead{(erg s$^{-1}$)} & \colhead{($1\sigma$ error)} & \colhead{(erg s$^{-1}$)} & \colhead{($1\sigma$ error)}
}
\decimalcolnumbers
\startdata
+3 & 61.92 & 11.94 & 8.46 & 1.69 \\
+129 & 47.75 & 3.40 & 0.91 & 0.23 \\
+187 & 24.20 & 1.57 & 0.77 & 0.10 \\
+246 & 30.21 & 1.89 & 0.78 & 0.09 \\
+293 & 18.90 & 1.20 & 1.10 & 0.09 \\
+505 & 11.76 & 0.75 & 0.83 & 0.07 \\
+570 & 12.33 & 1.08 & 0.40 & 0.16 \\
+684 & 8.36 & 0.92 & 0.52 & 0.19 \\
+1359 & 4.41 & 0.29 & 0.13 & 0.02 \\
\enddata

\end{deluxetable}

\begin{deluxetable*}{rcccccc}
\tablecaption{Blackbody fits to the spectral continua \label{tab:BB_fits}}
\tablewidth{0pt}
\tablehead{
\colhead{Phase} & \multicolumn2c{T$_{BB}$} & \multicolumn2c{$10^{41}$ L$_{BB}$} & \multicolumn2c{$10^{14}$ R$_{BB}$}\\
\colhead{(days)} & \colhead{(K)} & \colhead{($1\sigma$ error)} & \colhead{(erg s$^{-1}$ )} & \colhead{($1\sigma$ error)} & \colhead{(cm)} & \colhead{($1\sigma$ error)}
}
\decimalcolnumbers
\startdata
+3 & 8052 & 38 & 34.08 & 2.17 & 10.67 & 0.36 \\
+129 & 7432 & 50 & 5.67 & 0.38 & 5.11 & 0.19 \\
+187 & 7528 & 47 & 2.78 & 0.18 & 3.49 & 0.12 \\
+246 & 7543 & 41 & 4.19 & 0.27 & 4.26 & 0.15 \\
+293 & 8140 & 87 & 2.84 & 0.22 & 3.01 & 0.13 \\
+505 & 8815 & 90 & 2.24 & 0.17 & 2.28 & 0.10 \\
+570 & 9012 & 97 & 1.83 & 0.13 & 1.97 & 0.08 \\
+684 & 10170 & 146 & 1.99 & 0.16 & 1.62 & 0.08 \\
+1359 & 6942 & 58 & 0.65 & 0.04 & 1.99 & 0.08 \\
\enddata

\end{deluxetable*}

\subsection{Environment Analysis} \label{sec:host}

\subsubsection{MUSE Spectroscopy} \label{sec:muse_spec}

We use the MUSE data to investigate the HII regions and the environment around SN 2011fh. The datacube was corrected for reddening and redshift using the values from Section \ref{sec:host_gal}. To avoid contamination from the light of the SN, we masked a $6\arcsec$ circular region around SN 2011fh. We used {\tt\string ifuanalysis} \footnote{\url{https://ifuanal.readthedocs.io/en/latest/index.html}} to select and estimate the properties of the HII regions. {\tt\string ifuanalysis} uses {\tt\string STARLIGHT} \citep{2005MNRAS.358..363C, 2006MNRAS.370..721M, 2007MNRAS.381..263A} to fit the stellar continuum with stellar population synthesis models, and {\tt\string Astropy} \citep{astropy:2013, astropy:2018} to fit the emission lines with Gaussians. 

\begin{figure*}[t!]
\centerline{\includegraphics[scale=.46]{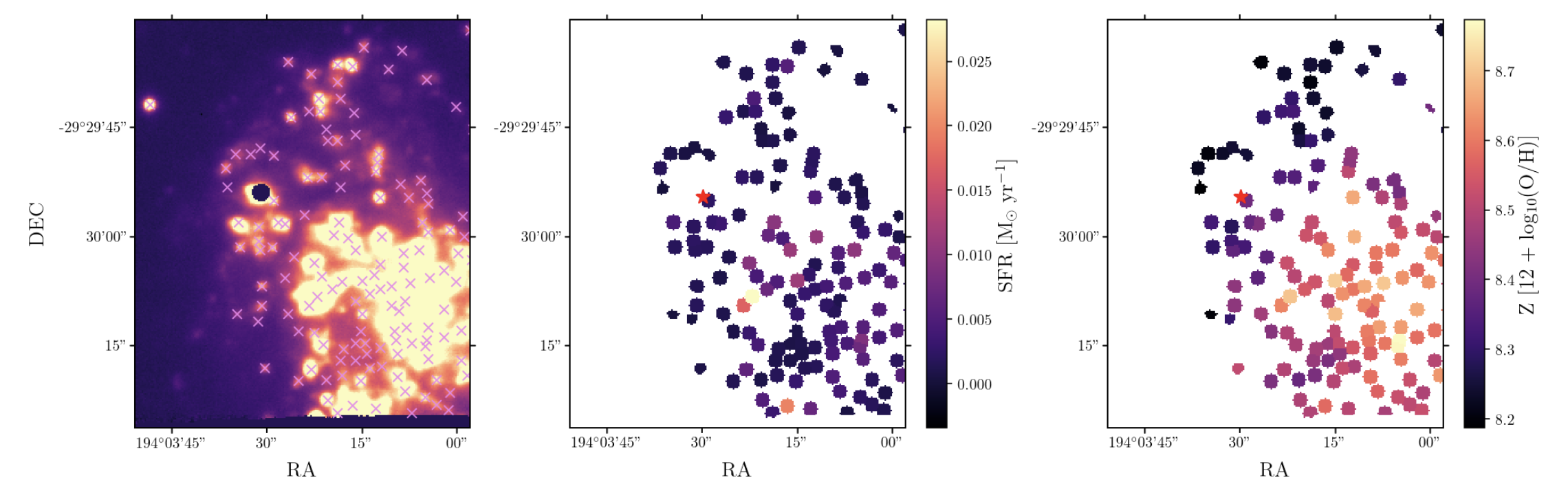}}
\caption{The left panel shows the HII regions selected by {\tt\string ifuanalysis} superimposed on the H$\alpha$ emission map. A $6\arcsec$ region around SN 2011fh was masked to avoid contamination from the SN. The middle panel shows the estimated star formation rate for these regions, in M$_\odot$ yr$^{-1}$, and the right panel shows the metallicity in units of $12 + \textrm{log}_{10}(\textrm{O/H})$. In the last two panels, the position of SN 2011fh is marked with a red star. \label{fig:ifu_z_sfr}}
\end{figure*}

The HII regions were selected to have a minimum peak H$\alpha$ flux twice the standard deviation of a region outside of the galaxy, and a maximum spatial size of 5 pixels.
In Figure \ref{fig:ifu_z_sfr}, we show the selected regions superimposed on the H$\alpha$ emission.
We also show the estimated oxygen abundances in units of $12 + \textrm{log}_{10}(\textrm{O/H})$, based on the calibration of \cite{2016Ap&SS.361...61D}, and the star formation rate (SFR), estimated using the H$\alpha$ luminosity as described in \cite{1998ARA&A..36..189K}, for each HII region. 
SN 2011fh is located in the outskirts of the galaxy, and the oxygen abundance value of its nearest HII region is $8.39 \pm 0.03$, or $0.5 \ \textrm{Z}_\odot$, assuming the Solar oxygen abundance of $12 + \textrm{log}_{10}(\textrm{O/H}) = 8.69$ \citep{2021arXiv210501661A}. This value is surprisingly similar to the metallicity estimated for SN 2009ip by \cite{2014ApJ...780...21M}.
The SFR is fairly constant throughout the galaxy, with a median value of $1.8 \times 10^{-3}$ M$_\odot$ yr$^{-1}$, although there is one HII region with a high value of $4.8 \times 10^{-2}$ M$_\odot$ yr$^{-1}$. SN 2011fh is located near a region with relatively high SFR of $\sim 2.3 \times 10^{-3}$ M$_\odot$ yr$^{-1}$.

\subsubsection{HST Analysis} \label{sec:HST}

As can be seen in Figure \ref{fig:fig_hst}, there are a number of bright sources within $\sim 150$ pc of SN 2011fh, probably forming an association with its progenitor star. We used  {\tt\string DOLPHOT} \citep{2016ascl.soft08013D} to perform PSF photometry on these sources, using the same selection criteria as described in Section \ref{sec:hst_imaging}. We selected a total of 10 stars within a $1\arcsec$ radius (150 pc) centered in SN 2011fh. The extinction corrected absolute magnitudes of the stars in the F336W and F814W filters are shown in Table \ref{tab:hst_cluster}.

We used {\tt\string HOKI} \citep{2020JOSS....5.1987S}, and its modules {\tt\string CMD} and {\tt\string AgeWizard} \citep[][]{2020MNRAS.498.1347S}, to estimate the age of the stellar cluster around SN 2011fh, and thus constrain the age of its progenitor. 
{\tt\string HOKI} uses stellar population models from Binary Population and Spectral Synthesis \citep[{\tt\string BPASS}, ][]{2017PASA...34...58E, 2018MNRAS.479...75S}.

\begin{figure}[h!]
\centerline{\includegraphics[scale=.5]{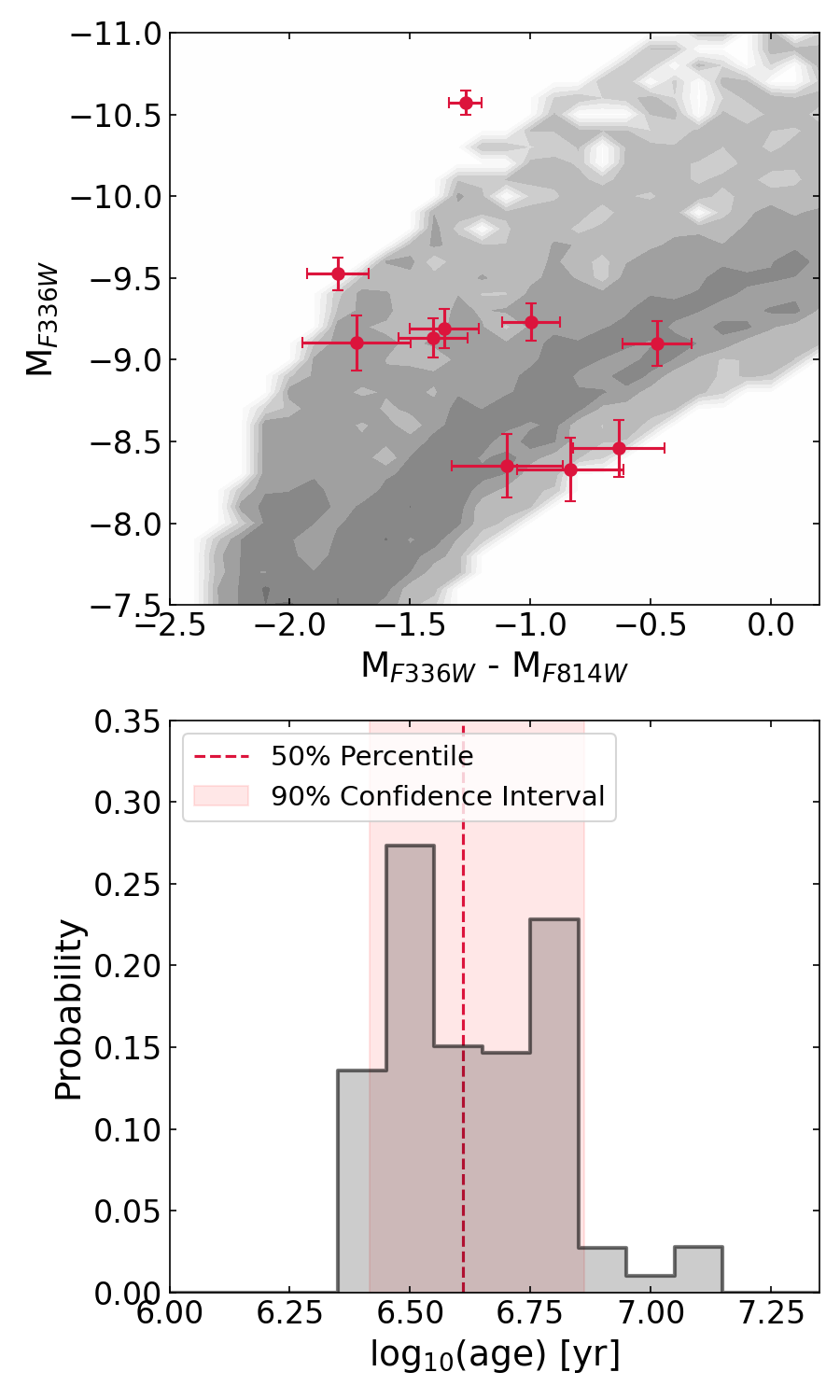}}
\caption{The top panel shows the color magnitude diagram (CMD) of the bright stars within $1 \arcsec$ (150 pc) of SN 2011fh, along with a {\tt\string HOKI} CMD with an age of $10^{6.7}$. The bottom panel shows the {\tt\string AgeWizard} probability distribution. The 90 percent confidence interval of the distribution is highlighted in the red region, and the median is marked as the dashed line. \label{fig:fig_agewiz}}
\end{figure}

The selected sources are very blue, which suggests a very young progenitor star for SN 2011fh. In Figure \ref{fig:fig_agewiz}, we show the $\textrm{M}_{F336W}$ versus $\textrm{M}_{F336W} - \textrm{M}_{F814W}$ color-magnitude diagram of the HST sources, along with the {\tt\string HOKI} synthetic CMD. Figure \ref{fig:fig_agewiz} shows that the detected sources are limited to a magnitude of $\textrm{M}_{F336W} \lesssim -8.4$~mag, which affects the detection of fainter sources and places an uncertainty in our analysis. We notice that the crowding of the stars in the region might also affects the detection of sources.
We set the metallicity to a value of $0.5 \ \textrm{Z}_\odot$ and found the best-fitting BPASS model to have an age of $10^{6.7}$ yr. The model was then used in {\tt\string AgeWizard} to estimate the probability distribution of ages based on the observed color-magnitude values. The resulting distribution is shown in the bottom panel of Figure \ref{fig:fig_agewiz}. The $90 \%$ confidence interval spans between $10^{6.4} - 10^{6.8}$ yr, with a median value of $10^{6.6}$ yr. This result indicates that the local stellar population, and therefore the progenitor of SN 2011fh, is very young.

\begin{deluxetable}{rcrc}
\tablecaption{HST photometry of the stars around SN 2011fh \label{tab:hst_cluster}}
\tablewidth{0pt}
\tablehead{
\multicolumn2c{F336W} & \multicolumn2c{F814W} \\
\colhead{(mag)} & \colhead{(error)} & \colhead{(mag)} & \colhead{(error)}\\
}
\decimalcolnumbers
\startdata
$-10.57$ & $0.10$ & $-9.30$ & $0.07$  \\
$-9.10$ & $0.16$ & $-8.63$ & $0.08$  \\
$-9.23$ & $0.13$ & $-8.23$ & $0.09$  \\
$-9.52$ & $0.12$ & $-7.73$ & $0.11$  \\
$-9.19$ & $0.14$ & $-7.83$ & $0.11$  \\
$-9.13$ & $0.14$ & $-7.73$ & $0.11$  \\
$-8.46$ & $0.19$ & $-7.83$ & $0.10$  \\
$-9.10$ & $0.19$ & $-7.38$ & $0.17$  \\
$-8.33$ & $0.21$ & $-7.49$ & $0.13$  \\
$-8.35$ & $0.21$ & $-7.26$ & $0.15$  \\
\enddata

\end{deluxetable}

\section{Discussion} \label{sec:dis}

The members of the family of SN 2009ip-like transients show remarkable similarities throughout their evolution, which may be indicative of similar progenitors or explosion mechanisms. In Section \ref{sec:relation_2009ip}, we describe the photometric and spectroscopic similarities of SN 2011fh to SN 2009ip and similar events. We discuss the possibility of dust formation in Section \ref{sec:presence_dust}. We analyze the late-time spectrum of SN 2011fh in Section \ref{sec:late_time}, where we also make a parallel with LBV eruptions. The implications of the age estimate for the progenitor star, such as the progenitor mass and possible explosion mechanisms, are discussed in Section \ref{sec:precursor}. 

\subsection{The relation with SN 2009ip-like events} \label{sec:relation_2009ip}

As described in Section \ref{sec:phot_evol} and shown in Figure \ref{fig:fig_compare_lc}, SN 2011fh, SN 2009ip \citep{2013MNRAS.430.1801M, 10.1093/mnras/stt813}, SN 2016bdu \citep{10.1093/mnras/stx2668}, and SN 2018cnf \citep{2019A&A...628A..93P} all show a similar brightening event (Event A) with M$_r \sim -15$ mag (M$_r \sim -16$ mag for SN 2011fh), followed by a luminous outburst (Event B) with M$_r \sim -18$ mag. SN 2009ip had a peak luminosity of $\sim 2 \times 10^{43}\ \textrm{erg s}^{-1}$  \citep{2013MNRAS.430.1801M,2014ApJ...780...21M}, while SN 2016bdu and SN 2018cnf peaked at $\sim 4 \times 10^{42}\ \textrm{erg s}^{-1}$ \citep{10.1093/mnras/stx2668, 2019A&A...628A..93P}, and SN 2011fh at $\sim 3.4 \times 10^{42}\ \textrm{erg s}^{-1}$ (Figure \ref{fig:fig_bb_superbol}). 

The interpretation of the precursor event and its connection to the brighter one is still unclear, with ideas ranging from an explosion (faint core-collapse or a major outburst in a massive star) followed by ejecta-CSM interaction \citep{2015ApJ...803L..26M, 2016MNRAS.463.3894E, 10.1093/mnras/stx2668}, an outburst preceding a SN \citep{2013Natur.494...65O, 10.1093/mnras/stx2668}, or the result of binary interactions \citep[][]{2013MNRAS.436.2484K, 2014AJ....147...23L, 2016AstBu..71..422G}. The slow brightening seen for SN 2011fh in the months before peak might indicate an LBV-like eruptive state shortly before a major outburst, as it is commonly observed for SN impostors \citep[][]{2011MNRAS.415..773S}, or even Type IIn SNe \citep[][]{2014ApJ...789..104O, 2021ApJ...907...99S}. The mechanism behind Event B, in this case, could be either the shock generated by a real core-collapse of the progenitor star, or some energetic eruption originating in instabilities that are thought to operate in massive stars, such as pulsational pair-instability \citep[][]{2007Natur.450..390W}, unsteady burning or interaction with other stars \citep[][]{2014ARA&A..52..487S, 2014ApJ...785...82S}. We cannot rule out the possibility that the first brightening was in fact the main explosive event, and that the following event was generated by its shock interacting with CSM shells. This scenario was invoked by \cite{10.1093/mnras/stx2668} in the case of SN 2016bdu, and by \cite{2016MNRAS.463.3894E} for SN 2015bh. \cite{2015ApJ...803L..26M} described a similar scenario in which the Event B observed in SN 2009ip was generated by shell-shell collisions. Unfortunately, the slow brightening event of SN 2011fh was only observed in $r$-band, and the lack of other colors or spectroscopic observations during this period makes it hard to constrain its true nature. 

Pre-discovery images of SN 2009ip and SN 2016bdu have shown LBV-like variability for their progenitors, with observed outbursts reaching M$_r \sim -14$ mag \citep[][]{2013MNRAS.430.1801M, 10.1093/mnras/stx2668}. In Figure \ref{fig:fig2} we have shown that nearly five years before the luminous outburst of 2011 the source ranged between M$_r \sim -14.5$ mag and M$_r \sim -15.5$ mag. 
We note, however, that these observations are relatively shallow, and these oscillations might not be real. For this reason, it is not possible to conclude if such behavior is related to a physical process in the progenitor star. 

In the days following peak brightness, SN 2009ip, SN 2016bdu, and SN 2018cnf all show a very similar ``shoulder'' in their light curve (see Figure \ref{fig:fig_compare_lc}), probably caused by a collision between CSM shells \citep[][]{2013MNRAS.430.1801M, 10.1093/mnras/stt813, 10.1093/mnras/stx2668, 2019A&A...628A..93P}. Although photometry of SN 2011fh is not available at this period, bumps in its $r$-band light curve are seen at $\sim 228$ and $\sim 350$ days after peak, probably generated by the same mechanism. The late-time brightness evolution of SN 2009ip and SN 2016bdu is very slow, probably due to ongoing interaction with a dense CSM \citep[][]{2017MNRAS.469.1559G, 10.1093/mnras/stx2668}. SN 2011fh also shows a slow evolution at later times, with an apparent flattening of the light curve after 2014. The 2016 October 20 HST observations revealed a source with an absolute magnitude of M$_{F814W} \approx -13.32$ at the position of SN 2011fh. 
This is more luminous than the expected for a star in a quiescent stage, which might suggest that the progenitor star is going through an LBV-like eruptive state. However, emission due to continuous interaction between the ejecta and the CSM is a common mechanism used to explain late-time emission of Type IIn SNe and other interacting transients \citep[][]{2013A&A...555A..10T, Fraser2020}, and might explain this late-time emission for SN 2011fh.

\begin{figure}[h!]
\centerline{\includegraphics[scale=.40]{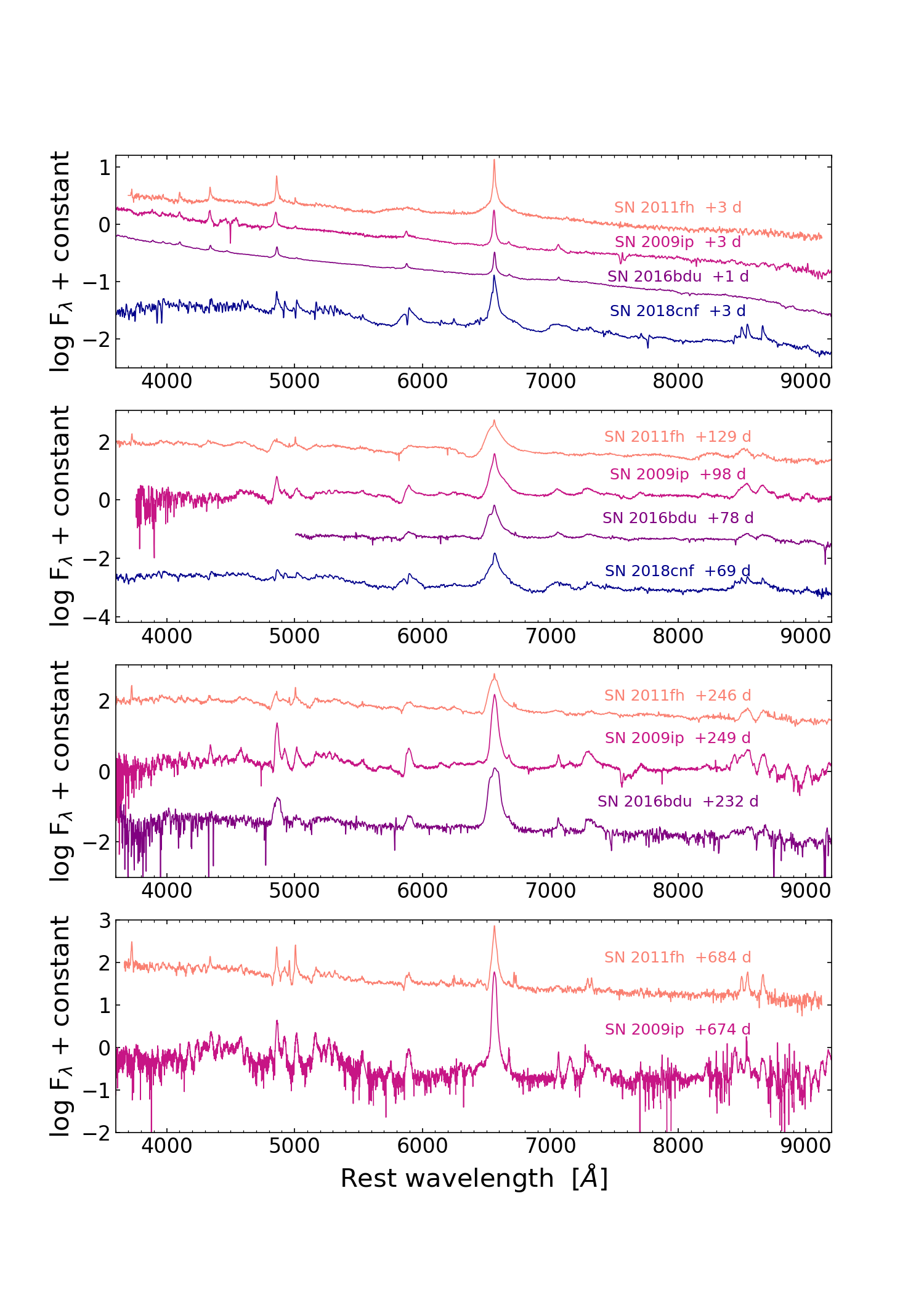}}
\caption{Spectral comparison of SN 2011fh to SN 2009ip \citep{2013MNRAS.433.1312F, 2014ApJ...780...21M, 2015MNRAS.453.3886F, 2017MNRAS.469.1559G}, SN 2016bdu \citep{10.1093/mnras/stx2668}, and SN 2018cnf \citep{2019A&A...628A..93P} at peak brightness and near phases $\sim$ 100, 240, and 680 days. \label{fig:comparison}}
\end{figure}

Besides their photometric correspondences, SN 2009ip-like events also show remarkable similarities in their spectroscopic evolution. In Figure \ref{fig:comparison}, we show the comparison of the spectral evolution of SN 2011fh to SN 2009ip \citep{2013MNRAS.433.1312F, 2014ApJ...780...21M, 2015MNRAS.453.3886F, 2017MNRAS.469.1559G}, SN 2016bdu \citep{10.1093/mnras/stx2668}, and SN 2018cnf \citep{2019A&A...628A..93P}. The four objects are very similar near peak brightness, with blue continua and prominent narrow to intermediate width Balmer emission features. SN 2011fh and SN 2018cnf, however, have lower blackbody temperatures, with a measured $\sim 8.3 \times 10^3$  K for the latter, while SN 2009ip and SN 2016bdu show temperatures of $\sim 1.2\times 10^4$ K and $\sim 1.7 \times 10^4$ K, respectively \citep{2014ApJ...787..163G, 10.1093/mnras/stx2668}. 

At $+129$ days, SN 2011fh has a very distinctive P-Cygni absorption feature, which indicate the presence of fast-moving material at $\sim 6400 \ \textrm{km s}^{-1}$. A similar feature with a velocity of $\sim 5 000 \ \textrm{km s}^{-1}$ was seen for SN 2009ip at similar epochs \citep{2014ApJ...780...21M}. At $+78$ days, SN 2016bdu also shows a prominent absorption feature at $\sim 3800 \ \textrm{km s}^{-1}$ \citep{10.1093/mnras/stx2668}. We note that higher velocity components were detected for both transients at earlier epochs, with measured FWHM values of up to $\sim 13000 \ \textrm{km s}^{-1}$ \citep{2014ApJ...780...21M, 10.1093/mnras/stx2668}. Another SN 2009ip-like event, LSQ13zm, presented even higher measured velocities of up to $\sim 22000 \ \textrm{km s}^{-1}$ \citep[][]{2016MNRAS.459.1039T}.

For spherical symmetry, it is possible to estimate the mass-loss rate of the progenitor star assuming that ejecta-CSM shock interaction powers the light curve \citep[e.g., ][]{2013MNRAS.429.2366S}. Using the CSM luminosity, $L_{CSM}$, the velocity of the cold dense shell, $V_{CDS}$, and the velocity of the unshocked CSM, $V_{CSM}$, and assuming 100\% efficiency of converting shock kinetic energy into radiation,

\begin{equation}
    \dot{M} = 2 \ L_{CSM} \ \frac{V_{CSM}}{V_{CDS}^3}.
\end{equation}

\noindent Using $V_{CDS} \approx 3000 \ \textrm{km} \ \textrm{s}^{-1}$, $V_{CSM} \approx 100$~km s$^{-1}$, and $L_{CSM} \simeq L{peak} \approx  3 \times 10^{42} \ \textrm{erg s}^{-1}$, we get $ \dot{M} \sim 4 \times 10^{-2} \ \textrm{M$_{\odot}$ yr}^{-1}$. This value is fully consistent with the estimated mass-loss rate for the progenitor of  SN 2009ip of $\sim 10^{-2} - 10^{-1} \ \textrm{M$_{\odot}$ yr}^{-1}$ \citep{2013MNRAS.433.1312F,2014ApJ...780...21M}. 
The observed mass-loss rate of SN 2011fh also falls within the observed range of mass-loss rates for Type IIn SNe of $10^{-2} - 10^{-1}\ \textrm{M$_{\odot}$ yr}^{-1}$ \citep[][]{2012ApJ...744...10K} and $10^{-4} - 10^{-2}\ \textrm{M$_{\odot}$ yr}^{-1}$ \citep[][]{2013A&A...555A..10T}.

As noted by \cite{2021arXiv210209572B}, 2009ip-like transients often show asymmetries in their H$\alpha$ emission profile. The line in AT 2016jbu was very asymmetric, in SN 2009ip it was very symmetric, and in SN 2016bdu it was somewhat asymmetric. SN 2011fh shows a degree of H$\alpha$ symmetry somewhat similar to SN 2016bdu, with a clearly asymmetrical shape until phase $+293$ (See Figure \ref{fig:fig_spec} and Figure \ref{fig:fig_Ha_fit}). This behavior of the H$\alpha$ profile might be related to asymmetries in the CSM

At late-times ($\sim 240$ days after peak), SN 2011fh, SN 2009ip, and SN 2016bdu are still dominated by emission lines of H and He, which indicates the ongoing interaction with a dense CSM. 
 All three objects show Ca II emission at this phase, and both SN 2009ip and SN 2016bdu show He I $\lambda$7065 emission, which will develop later in SN 2011fh.
The Fe emission line forest is also visible for the three events, and is particularly strong for SN 2009ip.
\cite{10.1093/mnras/stx2668} noted that SN 2016bdu shows nebular lines of [O I] at this epoch, and that those lines are expected for core-collapse SNe. The same was true for SN 2011fh. Although differing in some details, the H$\alpha$ profile of all the three events around this epoch can be described by a narrow feature superimposed on an intermediate width feature. 

As noted by \cite{2017MNRAS.469.1559G} for SN 2009ip, SN 2011fh also shows a distinctive O I $\lambda$8446 emission line but no O I $\lambda$7774 at very late-times. This suggests that O I $\lambda$8446 is being created by Bowen Ly-$\beta$ fluorescence, which is consistent with a very dense and optically thick medium.
\cite{2015MNRAS.453.3886F} showed that SN 2009ip had a ratio between the intensity of the [O I]$\lambda \lambda 6300, 6363$ lines of $\sim 1$. Although this ratio suggests newly synthesized oxygen in the ejecta, \cite{2015MNRAS.453.3886F} argue that the lack of evolution of this value points towards the presence of primordial oxygen. We note a very similar scenario for SN 2011fh, which has a ratio of $0.9 \pm 0.3$ at the phase of $+684$ days and $1.6\pm 0.3$ at $+1359$ days.
As seen in Figure \ref{fig:comparison}, SN 2011fh and SN 2009ip also show other similarities at phase $\sim +680$ days, such as the presence of strong Ca II features and the Fe II emission line forest. The $\textrm{H}\alpha/\textrm{H}\beta$ flux ratio for SN 2009ip of around 40 at this epoch is very high \citep{2017MNRAS.469.1559G}, while we estimate $\textrm{H}\alpha/\textrm{H}\beta \approx 7$ for SN 2011fh.

\subsection{The presence of dust} \label{sec:presence_dust}

In Section \ref{sec:sed}, we found that the spectral energy distribution (SED) of SN 2011fh has a significant IR excess that can be described by dust emission with a temperature of $\sim 950$~K at phase $+570$ days, and $\sim 1116$~K at phase $+1143$ days. 

The dust emission is located at $\sim 10^{16.52} \ \textrm{cm}$ at phase $+570$, and decreases to $\sim 10^{16.44} \ \textrm{cm}$ at later epochs (see Table \ref{tab:dusty_fits}).
Dust emission is quite common for interacting transients \citep[][]{2008ApJ...681L...9P, 2009ApJ...699.1850B, 2011ApJ...732...32F, 2011ApJ...741....7F}, but distinguishing between pre-existing or newly formed dust is not easy. SN 2011fh shows a similar mid-IR color evolution to several Type IIn SNe and other interacting transients, such as SN 2009ip \citep[][]{2021arXiv210612427S}. Asymmetries in the H$\alpha$ caused by the suppression of the red wing relative to the blue wing are often used as an indicator for the detection of newly formed dust \citep[e.g.,][]{2011ApJ...741....7F}. Although SN 2011fh has small asymmetries, it is hard to conclude whether it is caused by dust formed in the luminous outburst. The dust inner radius of $\sim 10^{16.5} \ \textrm{cm}$ is larger than the photospheric radius of $\sim 10^{14} \ \textrm{cm}$.
This indicates that dust is expanding slowly, if at all, and show little optical depth evolution, all of which suggests that dust is pre-existing.

The dust mass is

\begin{equation}
    M_{dust} = \frac{4 \pi \tau_V R_{dust}^2}{\kappa_V},
\end{equation}

\noindent where $\kappa_V \approx 2000 \ \textrm{cm}^2 / \textrm{g}$ is the dust visual opacity for silicate dust \citep{2010ApJ...725.1768F}. Using the properties estimated at $+570$ days, we get $\sim 2 \times 10^{-3} \ \textrm{M}_\odot$. For a typical gas to dust mass ratio of $\sim 100$, the total mass associated with the dust shell is of $\sim 0.2 \ \textrm{M}_\odot$. These values are much larger than were estimated for the LBV outburst UGC 2773 OT2009-1 \citep{2011ApJ...732...32F} and for the SN 2009ip-like event AT 2016jbu \citep[][]{2021arXiv210209572B, 2021arXiv210209576B}, but fall well within the range of dust masses estimated for Type IIn SNe by \cite{2011ApJ...741....7F}.

\subsection{The late-time spectrum} \label{sec:late_time}

Type IIn SNe often show spectra dominated by Balmer lines in emission until very late periods, instead of the nebular lines of heavy elements seen in non-interacting SNe. Some known exceptions are SN 1998S, which showed emission lines of [O I] and [Ca II] in the years after the explosion \citep[][]{2000ApJ...536..239L}, and SN 1995N, where nebular lines of [O I], [O II], [O III] appeared at very late times \citep[][]{2002ApJ...572..350F}. As was shown by \cite{2013A&A...555A..10T}, Type IIn SNe can show H$\alpha$ features with velocities of $\sim 10^{4} \ \textrm{km s}^{-1}$ and luminosities between $10^{40} - 10^{41} \ \textrm{erg s}^{-1}$ up to 180 days after peak. The spectra of SN impostors are also dominated by Balmer lines at late times, and many show Ca II emission lines \citep[][]{2011MNRAS.415..773S}.
One example is SN 2000ch, which reached M$_R \sim -12.8$ mag \citep[][]{2004PASP..116..326W} and had broad lines of H$\alpha$  with FWHM velocities of $\sim 1500 \ \textrm{km s}^{-1}$ \citep{2011MNRAS.415..773S}.

As described in Section \ref{sec:spec_evol}, the very late-time spectrum of SN 2011fh shows prominent lines of Fe II, Ca II, O I, and He I, while still being dominated by strong Balmer features. The H$\alpha$ profile at this epoch shows an intermediate-width component with a FWHM velocity of $\sim 726  \ \textrm{km s}^{-1}$, and a luminosity of $\sim 4 \times 10^{39} \ \textrm{erg s}^{-1}$. 

\begin{figure}[t!]
\centerline{\includegraphics[scale=.36]{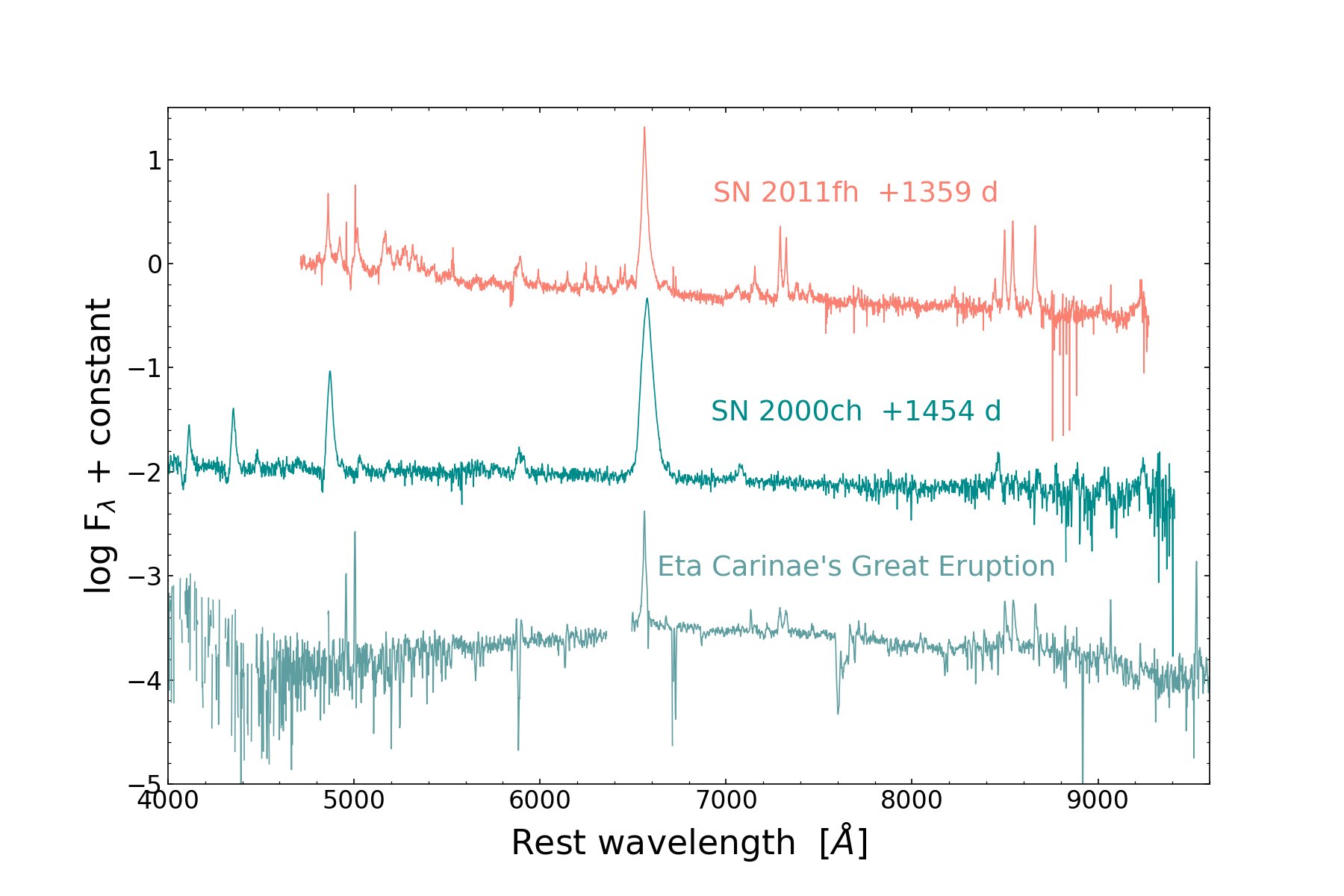}}
\caption{Spectral comparison of the very late-time spectrum of SN 2011fh, taken 1359 days after peak, to the spectrum of SN 2000ch at a similar phase \citep{2011MNRAS.415..773S}. We also show the light echo spectrum of the Great Eruption of Eta Carinae \citep{2018MNRAS.480.1466S}. \label{fig:comparison_impostors} }
\end{figure}

Although it is hard to determine whether SN 2011fh was a terminal event, we note several similarities between its late time spectrum and those of SN impostors. In Figure \ref{fig:comparison_impostors}, we compare the spectrum of SN 2011fh to SN 2000ch at phase $+1454$ days \citep{2011MNRAS.415..773S}, and to the 19th century Great Eruption of the Galactic LBV star Eta Carinae \citep{2018MNRAS.480.1466S}. SN 2011fh and SN 2000ch are both dominated by narrow to intermediate-width Balmer lines in emission until very late periods, with some He I features. The continuum slopes of the two spectra are very similar, indicating a similar photospheric temperature. SN 2000ch also shows a prominent O I $\lambda 8446$ emission line, which is also present in SN 2011fh. Although SN 2000ch does not show the strong Ca II features seen in SN 2011fh, these lines are present in the spectra of Eta Carinae's Great Eruption \citep{2018MNRAS.480.1466S}, and are strong features in the spectra of some events arising from lower mass progenitors, such as SN 2008S \citep{2009MNRAS.398.1041B, 2009ApJ...697L..49S}, NGC 300-OT \citep{2009ApJ...699.1850B, 2009ApJ...695L.154B}, and SN 2002bu \citep{2011MNRAS.415..773S}. We note that SN 2002bu and SN 2008S are sometimes referred to as intermediate-luminosity red transients (ILRTs), and that many studies support the scenario of SN 2008S-like events being terminal SN explosions \citep[][]{2016MNRAS.460.1645A, 2009MNRAS.398.1041B, 2021arXiv210805087C}. The late time spectrum of SN 2011fh also shows strong similarities with the spectra of UGC 2773 OT2009-1 \citep{2010AJ....139.1451S,2011ApJ...732...32F}, an analogue to Eta Carinae's Great Eruption \citep{2016MNRAS.455.3546S}. 

After peak, SN 2000ch remained in a plateau at $\sim -10.6$ mag, which could have been the quiescent magnitude of the progenitor star or a S-Dor like eruptive state \citep[][]{2004PASP..116..326W, 2011MNRAS.415..773S}. \cite{2010MNRAS.408..181P} described in detail other two bright eruptions related to the star, in 2008 and 2009, when it reached $\sim -12.8$ during the brightest one. \cite{2010MNRAS.408..181P} noted many observational similarities with the system HD 5980, which is a binary consisting of an erupting LBV and a Wolf-Rayet star, and suggested binary interactions as an explanation for the variability and eruptions detected for SN 2000ch. A similar scenario is often used to explain the Great Eruption of Eta Carinae.

\subsection{Implications of the progenitor age and mass} \label{sec:precursor}

\begin{figure}[t!]
\centerline{\includegraphics[scale=.7]{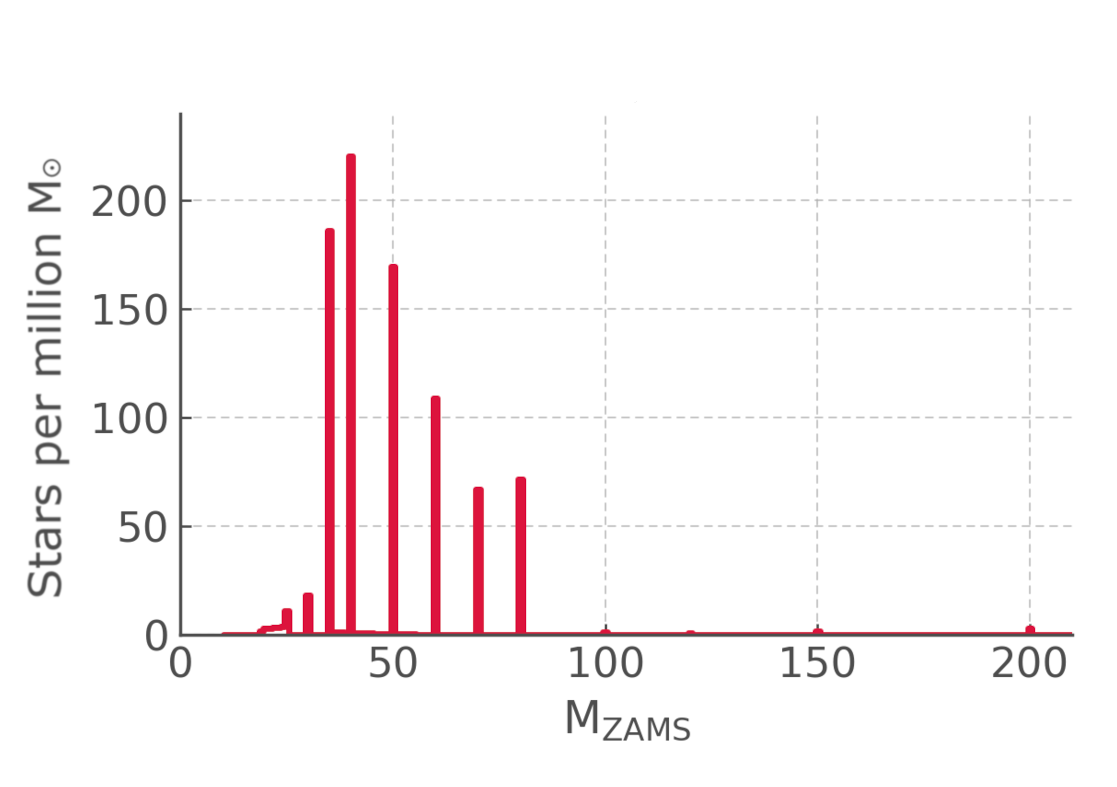}}
\caption{Number of stars per million M$_\odot$ obtained using {\tt\string BPASS}. The discrete number of ZAMS masses come from the grid of initial masses used in the {\tt\string BPASS} models. The highest bars in the distribution extend between 35 and 80 M$_\odot$.} \label{fig:bpass_ZAMS} 
\end{figure}

Our estimates of the age of the stars around SN 2011fh indicate that the progenitor is a very young star with $\sim 4.5$ Myr. Using $0.5 \ \textrm{Z}_\odot$ and a $4.5$~Myr {\tt \string MIST} \citep[][]{2011ApJS..192....3P, 2013ApJS..208....4P, 2015ApJS..220...15P, 2018ApJS..234...34P} single-star isochrone, we get a maximum mass of $\sim 60 \ \textrm{M}_\odot$. 
Since the fraction of binary systems is very high at large stellar masses, we also use {\tt\string BPASS} \citep[][]{2017PASA...34...58E, 2018MNRAS.479...75S} to estimate the progenitor mass. Using the age distribution of $2 - 6$~Myr (see Figure \ref{fig:fig_agewiz}), a half solar metallicity, and after selecting progenitors that could result in neutron star remnants with $<2.5 \ \textrm{M}_\odot$ \citep[based on Figure 3 of][]{PhysRevLett.121.161101} and cutting out pulsational pair-instability candidates that reached a He core mass of $>40 \ \textrm{M}_\odot$, we obtained the distribution of the number of stars per million solar masses as a function of their ZAMS mass, as shown in Figure \ref{fig:bpass_ZAMS}. The discrete number of ZAMS masses presented in Figure \ref{fig:bpass_ZAMS} is a consequence of the grid of initial masses used in the {\tt\string BPASS} models \citep[see][]{2017PASA...34...58E}. The distribution shows that there is a large number of stars populating the $35 - 80 \ \textrm{M}_\odot$ mass range, with the three tallest bars being between $35 - 50 \ \textrm{M}_\odot$. We note, however, that an energetic outburst such as SN 2011fh would be more likely to occur with stars in the most massive half of the distribution, even though the relative number of stars is slightly lower than the less massive half. The $\sim 60 \ \textrm{M}_\odot$, obtained by single-star models and the $35 - 80 \ \textrm{M}_\odot$ range, obtained considering binary systems are surprisingly similar to the estimates of $50 - 80 \ \textrm{M}_\odot$  \citep{2010AJ....139.1451S} and of $\gtrsim 60 \ \textrm{M}_\odot$ \citep{2011ApJ...732...32F} for SN 2009ip. It also agrees well with the results of \cite{2019A&A...628A..93P}, who showed that the progenitor of SN 2018cnf was consistent with a massive hypergiant or an LBV. We note that our results contrast with the estimates of \cite{2021arXiv210209572B, 2021arXiv210209576B}, who showed that the progenitor of AT 2016jbu was a relatively less massive hypergiant star with $\sim 20 \ \textrm{M}_\odot$. This might indicate that this type of event can be originated by progenitors with a large range of initial masses.

One common factor used to explain instabilities in LBV stars is their proximity to the classical Eddington limit, which would help to create strong winds. Eta Carinae is thought to have exceeded the Eddington limit by a factor of $\sim 5$ during its Great Eruption \citep[][]{2018MNRAS.480.1466S}. We estimate a still larger Eddington factor of $\sim 30$ for SN 2011fh at the beginning of the brightening event of February of 2011, when it had a magnitude of M$_r \approx -15.2$. An Eddington factor of $\sim 100$ is reached at the end of this period, just before the luminous outburst of August 2011. 

Super-Eddington winds can arise from energy deposited in the envelope of massive stars by different mechanisms, including wave heating and unstable fusion \citep{2012MNRAS.423L..92Q,2014ApJ...780...96S,2016MNRAS.458.1214Q}. Binary interactions may also be a mechanism for injecting energy into the outer layers of a massive star, since more than half of all massive stars are in binary systems with short enough periods to generate some kind of interaction or collision \citep{2012Sci...337..444S, 2014ARA&A..52..487S}. The interactions typically begin when the more massive star evolves off the main sequence. This is a scenario often used to explain the behavior of Eta Carinae  \citep[][]{2018MNRAS.480.1466S, 2011MNRAS.415.2009S}. 

The pulsational pair instability mechanism is capable of explaining the extreme mass loss seen in LBV stars \citep[][]{2007Natur.450..390W}, and was considered as an explanation for SN 2009ip by \cite{2013ApJ...767....1P}, \cite{2013MNRAS.430.1801M} and \cite{10.1093/mnras/stt813}. \cite{2014ApJ...780...21M}, however, demonstrated that, for a mass limit of 85 M$_\odot$ and a metallicity of $0.4 - 0.9$ Z$_\odot$, the progenitor of SN 2009ip would have a helium core mass lower than the $\sim 40$ M$_\odot$ needed for a pulsational pair instability. Since SN 2011fh is so similar to SN 2009ip, the pulsational pair instability seems as an unlikely mechanism for its bright eruptions.

Finally, we cannot fully exclude the possibility of SN 2011fh being a weak SN, although this appears less likely. Under-energetic explosions arising from core-collapse are expected to occur in such massive stars \citep[][]{2003ApJ...591..288H, 2010ApJ...719.1445M, 2016ApJ...821...38S}. \cite{2013MNRAS.430.1801M} used this mechanism to explain the double-peaked light curve of SN 2009ip, where Event A would be a weak core-collapse event, and Event B would be powered by the ejecta shock interacting with a dense CSM.
\cite{2014MNRAS.438.1191S} noted a similarity of the Event A of SN 2009ip to SN 1987A, a SN generated by a blue supergiant star \citep[][]{1989ARA&A..27..629A}. A similar statement was made by \cite{2016MNRAS.463.3894E} in the case of SN 2015bh. 
\cite{10.1093/mnras/stx2668} showed that other SN 2009ip-like events, such as SN 2016bdu, SN 2010mc \citep{2014MNRAS.438.1191S}, and LSQ13zm \citep[][]{2016MNRAS.459.1039T}, have Event A light curves that are very similar to the faint SN 1987A-like SN 2009E \citep[][]{2012A&A...537A.141P}. SN 2011fh, however, does not appear to present such similarities. 

\section{Conclusion} \label{sec:concl}

In this paper we presented optical, near-infrared and infrared observations of the Type IIn SN 2011fh, spanning from 2007 to 2017. The results can be summarized as follows:

\begin{itemize}
    \item SN 2011fh shows several similarities with the peculiar SN 2009ip and similar transients, such as SN 2016bdu and SN 2018cnf. They include: 1) an initial spectral resemblance to typical Type IIn SNe, with a blue continuum and narrow to intermediate width Balmer emission lines; 2) a brightening event that lasts $\sim 5$ months and reaches M$_r \sim -16$ mag, followed by a luminous outburst with M$_r \sim -18$ mag at peak; 3) a very similar light curve, with bumps and a slow fading with time; and 4) a very similar spectroscopic evolution, with signals of CSM interaction until very late times, distinctive P-Cygni features indicating a fast-moving material, and the presence of Ca II and Fe II lines. 
    
    \item The estimated progenitor mass-loss rate of $\sim 4 \times 10^{-2}$ M$_\odot$ yr$^{-1}$ is consistent with mass-loss rate estimates for SN 2009ip and Type IIn SNe. 
    
    \item SN 2011fh shows a significant IR excess that can be described by warm dust emission with $T_{dust} \approx 1000$~K and $R_{dust} \approx 3\times 10^{16}$~cm. The emission was probably created by pre-existing dust with a mass of $\textrm{M}_{dust} \sim 2 \times 10^{-3}$~M$_{\odot}$. 
    
    \item The very late-time spectrum of SN 2011fh shows strong similarities with some SN impostors observed at a same phase. The detection of narrow Ca II features and Balmer emission lines is also similar to the Great Eruption of Eta Carinae.
    
    \item SN 2011fh is located in a region with relatively rapid star formation ($\sim 2 \times 10^{-3}$ M$_\odot$ yr$^{-1}$). The H II region oxygen abundance of $12 + \textrm{log(O/H)} = 8.39$ ($\sim 0.5 \ \textrm{Z}_{\odot}$) is similar to what was estimated at the location of SN 2009ip.
    
    \item The local stellar population has an age of $\sim 4.5$~Myr, which corresponds to a progenitor with an initial main-sequence mass of $\sim 60$~M$\odot$, if we consider a single-stellar evolution model, or a range of $35 - 80 \ \textrm{M}_\odot$, considering binary systems. These mass estimates are similar to the estimates for SN 2009ip's progenitor, and consistent with very massive stars passing through the LBV phase.
    
    \item SN 2011fh exceeded the classical Eddington limit by a large factor in the months before the luminous outburst of 2011. This suggests that a possible mechanism behind the bright event is related to strong supper-Eddington winds.
    
\end{itemize}

Although SN 2011fh shows striking similarities with both impostors and 2009ip-like events, it is hard to conclude whether SN 2011fh was a genuine core-collapse SN, or a non-terminal event where the mass recently ejected collides with pre-existing CSM and generates a luminous outburst.
The lack of observational signatures related to a core-collapse, such as broad Balmer and O I lines at late times, makes the scenario of a true SN less likely. Our results therefore imply that at least a fraction of SN 2009ip-like events arise from non-terminal eruptions in massive and young stars. 
The discovery and follow up observations of new transients similar to SN 2009ip will certainly help in the understanding of the mechanisms behind this unique class of events. 


\vspace{1cm}

We would like to thank Andrea Pastorello and Melissa Graham for kindly sharing their data on SN 2016bdu, SN 2018cnf, and SN 2009ip. We thank the Carnegie Supernova Project-II for obtaining several photometric observations presented in the paper, and in particular Carlos Contreras, Mark Phillips, Nidia Morrell and Eric Hsiao. TP is supported by CONICYT's Programa de Astronom\'ia through the ALMA-CONICYT 2019 grant 31190017. Support for JLP is provided in part by ANID through the Fondecyt regular grant 1191038 and through the Millennium Science Initiative grant ICN12\underline009, awarded to The Millennium Institute of Astrophysics, MAS. CSK is supported by NSF grants AST-1814440 and AST-1907570. HFS acknowledges the support of the Marsden Fund Council managed through Royal Society Te Aparangi. This paper includes optical and NIR photometry obtained by the Carnegie Supernova Project, which was generously supported by NSF grants AST-1008343, AST-1613426, AST-1613455, and AST-1613472.

\software{
HOKI \citep{2020JOSS....5.1987S,2020MNRAS.498.1347S}, 
BPASS \citep[][]{2017PASA...34...58E, 2018MNRAS.479...75S},
FIREHOSE \citep[][]{2013PASP..125..270S}, 
IRAF \citep[][]{1986SPIE..627..733T, 1993ASPC...52..173T}, 
Astropy \citep{astropy:2013, astropy:2018}, Photutils \citep[][]{2019zndo...3568287B}, 
PYPHOT (\url{https://mfouesneau.github.io/docs/pyphot/}),
DUSTY \citep{1997MNRAS.287..799I}, 
DOLPHOT \citep{2016ascl.soft08013D}, 
SUPERBOL \citep[Version 1.7;][]{2018RNAAS...2..230N},
ifuanalysis (\url{https://ifuanal.readthedocs.io/en/latest/index.html}),
STARLIGHT \citep{2005MNRAS.358..363C, 2006MNRAS.370..721M, 2007MNRAS.381..263A}
}

\newpage

\bibliography{ref}
\bibliographystyle{aasjournal}

\appendix

\section{Contamination} \label{sec:contam}

SN 2011fh is located in a relatively crowded region of the NGC 4806 galaxy, with several bright stars in the proximity (see Figure \ref{fig:fig_hst}). These stars lie within 150 pc ($1\arcsec$) around SN 2011fh, with several brighter sources within 300 pc ($2\arcsec$). Since the pixel scale of the pre SN observations were limited up to $2\arcsec$, we decided to estimate the possible contamination of the stellar cluster to our photometric measurements.  

To quantify the amount of contamination in the optical bands, we first performed aperture photometry of the stars around SN 2011fh in the HST F336W and F814W images, using the {\tt\string Astropy} \citep{astropy:2013, astropy:2018} package {\tt\string Photutils} \citep[][]{2019zndo...3568287B} in a $2\arcsec$ region. After subtracting the flux of SN 2011fh, we obtained
a contribution from the stars in this aperture of 20.67 and 20.05 mag, in the F336W and F814W filters, respectively.
By assuming a flat spectrum for the stellar cluster, we then interpolated between the F336W and F814W filter fluxes and performed synthetic photometry in the resulting line with {\tt\string PYPHOT}\footnote{ https://mfouesneau.github.io/docs/pyphot/}, to obtain the estimates of $V \approx 20.17$, $r \approx 20.18$, and $i \approx 20.18$. If we compare these values with the PSF photometry of SN 2011fh presented in Tables \ref{tab:r-crts} and \ref{tab:BVri-csp2}, the estimated contribution from the cluster is significantly lower than the flux of SN 2011fh throughout its photometric evolution. ]

\section{Tables}

\startlongtable
\begin{deluxetable}{llcc}
\tablecaption{Amateur astronomers data and CRTS $r$-band photometry of SN 2011fh \label{tab:r-crts}}
\tablewidth{0pt}
\tablehead{
\colhead{UT Date} & \colhead{JD} & \multicolumn2c{r}\\
\colhead{}& \colhead{} & \colhead{(mag)} & \colhead{(error)}
}
\decimalcolnumbers
\startdata
2007-1-25 & 2454126.20298 & 17.879 & 0.12 \\
2007-2-22 & 2454154.20024 & 17.988 & 0.13 \\
2007-3-9 & 2454169.17404 & 18.272 & 0.178 \\
2007-3-27 & 2454187.19841 & 17.961 & 0.125 \\
2007-4-6 & 2454197.17815 & 18.547 & 0.25 \\
2007-4-16 & 2454207.0734 & 18.184 & 0.16 \\
2007-5-7 & 2454228.04447 & 18.278 & 0.177 \\
2007-5-14 & 2454235.05647 & 18.097 & 0.143 \\
2007-6-16 & 2454267.97322 & 17.943 & 0.131 \\
2007-7-4 & 2454285.85462 & 17.999 & 0.134 \\
2007-7-20 & 2454301.85781 & 18.232 & 0.166 \\
2007-8-12 & 2454324.88361 & 17.985 & 0.129 \\
2007-12-28 & 2454463.18063 & 18.09 & 0.136 \\
2008-1-22 & 2454488.1978 & 18.273 & 0.182 \\
2008-2-7 & 2454504.25914 & 18.057 & 0.135 \\
2008-2-8 & 2454505.25765 & 18.121 & 0.148 \\
2008-2-29 & 2454526.17647 & 17.988 & 0.135 \\
2008-3-11 & 2454537.18612 & 18.286 & 0.179 \\
2008-5-6 & 2454593.06685 & 17.937 & 0.127 \\
2008-5-11 & 2454598.10991 & 17.912 & 0.131 \\
2008-5-28 & 2454615.0704 & 18.239 & 0.179 \\
2008-6-27 & 2454644.9809 & 17.992 & 0.14 \\
2008-12-31 & 2454832.20995 & 18.118 & 0.15 \\
2009-1-5 & 2454837.49042 & 18.121 & 0.243 \\
2009-2-25 & 2454888.21891 & 18.207 & 0.165 \\
2009-3-15 & 2454906.27249 & 17.734 & 0.106 \\
2009-3-27 & 2454918.21621 & 17.877 & 0.127 \\
2009-4-15 & 2454937.1613 & 18.235 & 0.172 \\
2009-4-27 & 2454949.13822 & 17.857 & 0.11 \\
2009-4-29 & 2454951.09237 & 17.945 & 0.138 \\
2009-5-11 & 2454963.0813 & 18.071 & 0.156 \\
2009-5-12 & 2454964.30533 & 17.799 & 0.163 \\
2009-5-29 & 2454981.11436 & 17.79 & 0.126 \\
2009-6-19 & 2455002.00408 & 18.2 & 0.164 \\
2009-7-18 & 2455030.93214 & 18.241 & 0.166 \\
2010-1-21 & 2455218.17694 & 17.777 & 0.101 \\
2010-2-21 & 2455249.13263 & 18.289 & 0.171 \\
2010-3-9 & 2455265.12761 & 17.893 & 0.118 \\
2010-4-8 & 2455295.17376 & 18.055 & 0.131 \\
2010-4-17 & 2455304.11246 & 18.077 & 0.144 \\
2010-5-10 & 2455327.01388 & 17.666 & 0.091 \\
2010-5-22 & 2455339.01684 & 17.92 & 0.118 \\
2010-6-13 & 2455361.0242 & 17.762 & 0.107 \\
2010-6-30 & 2455378.24321 & 17.948 & 0.189 \\
2010-7-16 & 2455393.9443 & 17.752 & 0.118 \\
2010-7-22 & 2455399.93913 & 17.83 & 0.112 \\
2010-7-28 & 2455406.20502 & 17.969 & 0.2 \\
2011-2-11 & 2455603.5911 & 17.636 & 0.146 \\
2011-2-24 & 2455617.25406 & 17.216 & 0.061 \\
2011-2-24 & 2455617.4946 & 17.549 & 0.102 \\
2011-4-13 & 2455665.05019 & 17.037 & 0.054 \\
2011-4-24 & 2455676.10863 & 17.176 & 0.063 \\
2011-4-25 & 2455677.4096 & 17.271 & 0.098 \\
2011-5-9 & 2455691.0465 & 16.845 & 0.052 \\
2011-5-26 & 2455708.02969 & 16.936 & 0.053 \\
2011-6-2 & 2455715.3151 & 16.716 & 0.075 \\
2011-6-3 & 2455716.22604 & 16.883 & 0.072 \\
2011-6-16 & 2455728.86432 & 16.855 & 0.089 \\
2011-6-21 & 2455734.03201 & 16.671 & 0.073 \\
2011-6-27 & 2455739.96366 & 16.694 & 0.045 \\
2011-6-28 & 2455741.29443 & 16.799 & 0.067 \\
2011-6-30 & 2455742.91587 & 16.861 & 0.057 \\
2011-6-30 & 2455742.91587 & 16.837 & 0.058 \\
2011-7-17 & 2455759.84951 & 16.614 & 0.048 \\
2011-7-28 & 2455770.90473 & 16.264 & 0.036 \\
2011-8-2 & 2455776.20691 & 16.342 & 0.068 \\
2011-8-9 & 2455782.84013 & 15.726 & 0.037 \\
2011-8-24 & 2455798.24181 & 14.942 & 0.028 \\
2011-8-25 & 2455799.24933 & 14.962 & 0.029 \\
2011-9-4 & 2455809.21792 & 15.362 & 0.031 \\
2011-9-5 & 2455810.22498 & 15.321 & 0.033 \\
2011-12-16 & 2455911.57505 & 16.731 & 0.046 \\
2011-12-25 & 2455920.58029 & 16.704 & 0.041 \\
2012-1-6 & 2455932.58682 & 16.748 & 0.041 \\
2012-1-25 & 2455951.5711 & 17.092 & 0.051 \\
2012-1-28 & 2455954.56738 & 17.373 & 0.06 \\
2012-2-16 & 2455973.54651 & 17.275 & 0.061 \\
2012-2-17 & 2455975.08906 & 17.701 & 0.151 \\
2012-2-24 & 2455981.55593 & 17.543 & 0.078 \\
2012-3-25 & 2456012.45263 & 17.469 & 0.068 \\
2012-3-30 & 2456017.15598 & 17.379 & 0.073 \\
2012-4-10 & 2456028.43419 & 16.886 & 0.047 \\
2012-4-11 & 2456029.14365 & 16.912 & 0.055 \\
2012-4-19 & 2456037.0854 & 17.121 & 0.062 \\
2012-5-15 & 2456063.026 & 17.33 & 0.073 \\
2012-7-8 & 2456116.95661 & 17.984 & 0.21 \\
2012-8-12 & 2456151.87688 & 17.671 & 0.104 \\
2012-8-12 & 2456152.21301 & 17.248 & 0.063 \\
2012-12-17 & 2456278.54791 & 18.305 & 0.173 \\
2012-12-21 & 2456282.56931 & 18.225 & 0.167 \\
2013-1-1 & 2456294.20582 & 18.136 & 0.154 \\
2013-1-22 & 2456314.55775 & 18.103 & 0.183 \\
2013-2-7 & 2456330.57937 & 17.952 & 0.106 \\
2013-2-18 & 2456341.58446 & 18.258 & 0.161 \\
2013-2-21 & 2456344.5748 & 18.055 & 0.124 \\
2013-2-24 & 2456348.40002 & 18.19 & 0.236 \\
2013-3-2 & 2456354.49434 & 18.059 & 0.203 \\
2013-3-9 & 2456361.45919 & 17.904 & 0.164 \\
2013-3-17 & 2456368.53775 & 18.08 & 0.164 \\
2013-4-11 & 2456394.39371 & 18.169 & 0.142 \\
2013-4-19 & 2456402.41645 & 17.842 & 0.106 \\
2013-4-30 & 2456413.32136 & 18.037 & 0.134 \\
2013-5-1 & 2456414.3878 & 18.168 & 0.167 \\
2013-5-2 & 2456415.27285 & 18.079 & 0.152 \\
2013-5-4 & 2456417.32889 & 18.428 & 0.221 \\
2013-5-5 & 2456418.36387 & 18.204 & 0.146 \\
2013-5-10 & 2456423.27335 & 17.848 & 0.117 \\
2013-5-11 & 2456424.21595 & 18.38 & 0.201 \\
2013-5-12 & 2456425.26083 & 17.842 & 0.098 \\
2013-5-14 & 2456427.23059 & 18.231 & 0.178 \\
2013-5-16 & 2456429.43702 & 18.163 & 0.146 \\
2013-5-31 & 2456444.31843 & 18.282 & 0.171 \\
2013-6-3 & 2456447.31712 & 18.319 & 0.173 \\
2013-6-5 & 2456449.28426 & 18.329 & 0.169 \\
2013-6-15 & 2456459.21472 & 18.617 & 0.289 \\
2013-6-16 & 2456460.28568 & 18.334 & 0.193 \\
2013-6-26 & 2456470.25602 & 18.18 & 0.15 \\
2013-6-28 & 2456472.28668 & 18.422 & 0.194 \\
2013-6-29 & 2456473.2751 & 17.967 & 0.108 \\
\enddata

\end{deluxetable} 

\begin{deluxetable*}{llcccccccc}
\tablecaption{$BVri$-CSP2 photometry of SN 2011fh \label{tab:BVri-csp2}}
\tablewidth{0pt}
\tablehead{
\colhead{UT Date} & \colhead{JD} & \multicolumn2c{B} & \multicolumn2c{V} & \multicolumn2c{r}& \multicolumn2c{i} \\
\colhead{}& \colhead{} & \colhead{(mag)} & \colhead{(error)}& \colhead{(mag)} & \colhead{(error)}& \colhead{(mag)} & \colhead{(error)}& \colhead{(mag)} & \colhead{(error)}
}
\decimalcolnumbers
\startdata
2013-2-20 & 2456343.85 & 18.843 & 0.0215 & 18.493 & 0.0173 & 18.063 & 0.0132 & 18.4135 & 0.0151\\
2013-3-3 & 2456354.78 & 18.8695 & 0.0226 & 18.5055 & 0.0193& 18.0805 & 0.0142& 18.41 & 0.0185\\
2013-3-15 & 2456366.71& 18.9345 & 0.0244& 18.576 & 0.0193& 18.1375 & 0.0154& 18.5 & 0.0202 \\
2013-3-18 & 2456369.91 &-&-&-&- & 18.15 & 0.0155 & 18.407 & 0.0324\\
2013-3-25 & 2456376.76 &-&-&-&- & 18.155 & 0.0146  & 18.529 & 0.0213\\
2013-4-5 & 2456387.68 & 18.913 & 0.0193& 18.573 & 0.0195& 18.1565 & 0.0132& 18.525 & 0.0171\\
2013-4-20 & 2456402.73 & 18.962 & 0.0265& 18.5725 & 0.0244& 18.193 & 0.0182& 18.522 & 0.0324\\
2013-5-4 & 2456416.68 & 19.1855 & 0.0273 & 18.8505 & 0.0245& 18.1565 & 0.0132& 18.816 & 0.0232\\
2013-5-12 & 2456424.75 & 19.103 & 0.0233& 18.795 & 0.0243 & 18.3765 & 0.0151& 18.795 & 0.0251\\
2013-6-1 & 2456444.68& 19.046 & 0.0254&-&-& 18.341 & 0.0192& 18.717 & 0.0291 \\
2015-3-16 & 2457097.76 & 19.7675 & 0.0205& 19.559 & 0.0165& 18.341 & 0.0192 & 19.558 & 0.0177\\
2015-3-18 & 2457099.80 & 19.883 & 0.0266& 19.5225 & 0.0167 & - & -& - & -\\
2016-1-14 & 2457401.80 & 19.625 & 0.0341& 19.3705 & 0.0306& 19.229 & 0.0342& 19.475 & 0.03815\\ 
\enddata

\end{deluxetable*}

\begin{deluxetable*}{llcccccc}
\tablecaption{$YJH$-CSP2 photometry of SN 2011fh \label{tab:YJH-csp2}}
\tablewidth{0pt}
\tablehead{
\colhead{UT Date} & \colhead{JD} & \multicolumn2c{Y}& \multicolumn2c{J}& \multicolumn2c{H} \\
\colhead{}& \colhead{} & \colhead{(mag)} & \colhead{(error)}& \colhead{(mag)} & \colhead{(error)}& \colhead{(mag)} & \colhead{(error)}
}
\decimalcolnumbers
\startdata
2013-2-23 & 2456346.86 & 17.62 & 0.031 & - & - & - & - \\
2013-2-27 & 2456350.81 & 17.58 & 0.03 & 17.457 & 0.026 & 17.19 & 0.037\\
2013-3-26 & 2456377.75 & 17.75 & 0.023& 17.649 & 0.021 &-&-\\
2015-4-30 & 2457142.58 & 18.865 & 0.044& 18.833 & 0.051 & 18.491 & 0.0788\\
2016-3-27 & 2457474.70 & 18.575 & 0.075 & 18.64 & 0.085& 18.25 & 0.113 \\
2017-5-16 & 2457889.50 & 18.208 & 0.043 & - & -& - & - \\
\enddata

\end{deluxetable*}

\begin{deluxetable*}{lccccc}
\tablecaption{Spitzer photometry of SN 2011fh\label{tab:spit}}
\tablewidth{0pt}
\tablehead{
\colhead{UT Date} & \colhead{JD} & \multicolumn2c{Channel 1 (3.6 $\mu$m)} & \multicolumn2c{Channel 2 (4.5 $\mu$m)}\\
\colhead{} & \colhead{} & \colhead{(mag)} & \colhead{(error)} & \colhead{(mag)} & \colhead{(error)}
}
\decimalcolnumbers
\startdata
2009-8-6 & 2455050.094 & 16.019 & 0.047 & 15.686 & 0.045 \\
2009-9-5 & 2455080.09 & 16.142 & 0.043 & 15.77 & 0.047 \\
2013-3-31 & 2456383.113 & 15.899 & 0.045 & 15.024 & 0.04 \\
2014-4-4 & 2456752.49 & 16.27 & 0.048 & 15.572 & 0.04 \\
2014-9-12 & 2456913.073 & 16.228 & 0.059 & 15.492 & 0.045 \\
2014-10-12 & 2456943.206 & 16.24 & 0.052 & 15.529 & 0.044 \\
2015-5-16 & 2457158.625 & 16.265 & 0.052 & 15.64 & 0.045 \\
\enddata

\end{deluxetable*}

\end{document}